%% file: main.tex
\documentclass[journal,comsoc]{IEEEtran}
\usepackage[T1]{fontenc}

\usepackage{cite}

\ifCLASSINFOpdf
  \usepackage[pdftex]{graphicx}
  \graphicspath{{fig/}}
  \DeclareGraphicsExtensions{.pdf,.jpeg,.png}
\else
  \usepackage[dvips]{graphicx}
  \graphicspath{{fig/}}
  \DeclareGraphicsExtensions{.eps}
\fi

\usepackage[TABBOTCAP]{subfigure}

\usepackage{amssymb}
\usepackage{amsmath}
\usepackage[cmintegrals]{newtxmath}
\usepackage{bm}

\usepackage{algorithm}
\usepackage{algorithmicx}
\usepackage{algpseudocode}

  \newcommand{\NULL}{\textbf{null}}

\usepackage{color}
\newcommand{\comm}[1]{\hfill\textcolor[rgb]{0.3,0.3,0.3}{\# #1}}

\usepackage{array}
\usepackage{booktabs} 

\usepackage{url}

\begin{document}
%
\title{RVH: Range-Vector Hash for Fast Online \\ Packet Classification}

\author{Tong~Shen,
         Gaogang~Xie,
         Xin~Wang,
        Zhenyu~Li,
         Xinyi~Zhang,
         Penghao~Zhang,
      	Dafang~Zhang
}

%



\maketitle

\input{sect-abst}

%
\IEEEpeerreviewmaketitle

\input{sect-intro}

\input{sect-relat}

\input{sect-rvss}

\input{sect-model}

\input{sect-part}
\input{sect-exper}

\input{sect-discu}
\input{sect-conclu}


%



\input{sect-acks}

\ifCLASSOPTIONcaptionsoff
  \newpage
\fi



%


\bibliographystyle{IEEEtran}
\bibliography{bibliography}

%








\end{document}

%% file: sect-abst.tex

\begin{abstract}
Packet classification according to multi-field ruleset is a key component for many network applications. Emerging software defined networking and cloud computing need to update the rulesets frequently for flexible policy configuration. Their success depends on the availability of the new generation of classifiers that can support both fast ruleset updating and high-speed packet classification. However, existing packet classification approaches focus either on high-speed packet classification or fast rule update, but no known scheme meets both requirements.
In this paper, we propose \textit{Range-vector Hash} (RVH) to effectively accelerate the packet classification with a hash-based algorithm while ensuring the fast rule update. RVH is built on our key observation that the number of distinct combinations of each field prefix lengths is not evenly distributed.
To reduce the number of hash tables for fast classification, we introduce a novel concept \textit{range-vector} with each specified the length range of each field prefix of the projected rules.
RVH can overcome the major obstacle that hinders hash-based packet classification by balancing the number of hash tables and the probability of hash collision.
Experimental results demonstrate that RVH can achieve the classification speed up to 15.7 times and the update speed up to 2.3 times that of the state-of-the-art algorithms on average, while only consuming 44\% less memory.
\end{abstract}


%% file: sect-intro.tex

\section{Introduction}
\label{sect:intro}

\IEEEPARstart{P}{acket} classification, where each incoming packet is matched against a multi-field ruleset in classifier, is one of the essential operations applied in switches, routers and other network appliances to support security~\cite{suh2014building}, quality of service (QoS)~\cite{lenzen2016tight, seddiki2014flowqos} and advanced functions~\cite{hawilo2014nfv,grimes2004mobile}. For example, to protect network resources from being abused, five-field firewall rules are often added to a switch to determine if packets should be forwarded or dropped. Generally, in traditional applications, rules are kept relatively static, thus fast classification can be achieved with the algorithms running over well-designed data structure of classifiers built offline. 
In the past, the main goal of classifier design is to perform high-speed packet processing, such as content detecting, load balancing, packet filtering, and forwarding, within reasonable memory footprint. As the rule update is infrequent at that time, the classifier can be built offline.

The emergence of software defined networking (SDN)~\cite{Csikor2016Dataplane,Sivaraman2016Packet,Bojie2016ClickNP} creates enormous opportunities for network innovation to support new features and value-added functions. 
This includes the support of traffic engineering~\cite{jain2013b4}, network function virtualization (NFV)~\cite{Sun2017NFP,Rexford2017Dynamic} and high-performance cloud computing~\cite{Daniel2017VFP,Chang2016Embark}.
However, these new functions rely on the capability of dynamic update of rules in the classifier~\cite{yingchareonthawornchai2016sorted, MPD2015SDNFlowTable}, besides the basic fast packet classification. 
On the one hand, these new applications should respond to a large amount of requests from different users instantly, so that the classifier rules have to be updated frequently to meet different requirements. On the other hand, the regular migration of network functions and the alteration of network topology and policy will update the classifier rules accordingly. This requires packet classification to be performed online with the support of fast dynamic rule update, and which is a definitely mandatory requirement for current and future classifiers.

Although packet classification is very important and has drawn a lot of research attention, existing algorithms often cannot meet the above two requirements at the same time.
Decision-tree based algorithms, such as HyperCuts (HyC)~\cite{singh2003packet}, EffiCuts (EC)~\cite{vamanan2010efficuts} and SmartSplit (SS)~\cite{He2014Meta} can achieve high-speed classification but not fast rule update. Hash based algorithms, such as Tuple Space Search (TSS)~\cite{srinivasan1999packet} implemented in Open vSwitch (OVS)~\cite{pfaff2015design}, can achieve fast rule updating but not high-speed classification. PartitionSort (PS)~\cite{yingchareonthawornchai2016sorted} and TupleMerge (TM) \cite{Daly2017TupleMerge} can improve classification speed at the cost of updating time. Achieving fast packet classification and rule update simultaneously is one of the fundamental challenges that need to be addressed to meet the emerging needs of advanced network management and efficient cloud computing.

To support both high-speed packet classification and fast rule updates, we propose an online packet classification algorithm \textit{Range-vector Hash}~(RVH), which achieves ultra-performance rule update by using efficient hash function, and trades off between the number of hash tables and the probability of hash conflicts to minimize the packet classification time. Specifically, to reduce the classification time, we introduce a novel concept \textit{range-vector}. To reduce the number of hash tables thus the classification time, we partition the ruleset into a series of sub-rulesets which are covered by range-vectors based on their distribution of each field prefix length. Each range-vector maintains a hash table, and the classification rules are mapped to different range-vectors (hash tables) with a proper design to bring down hash conflicts. In this way, updating a rule requires only one hash operation on average. Our main contributions are summarized as follows:
\begin{itemize}
 \item We propose a novel concept range-vector to efficiently partition the rule space for both fast packet classification and rule update. RVH can largely reduce the hash operations for packet classification while only requiring one hash operation for a rule update.
 \item We propose the use of base-vector to guide the construction of hash keys in the presence of field prefixes with heterogenous lengths in the same hash table.
 \item We propose an analytical model to effectively capture the cost for packet classification, which can serve as a base to minimize the classification time with the proper selection of the number of hash tables to construct.
 \item We propose a partitioning policy which effectively reduces the overlap of rules thus additional overhead (e.g., hash conflict) during the classification.
\end{itemize}

The packet classification module with mega-flow and micro-flow in OVS is originally implemented with TSS~\cite{srinivasan1999packet}, and has been implemented with RVH, PS~\cite{yingchareonthawornchai2016sorted} and TM~\cite{Daly2017TupleMerge} in our experiments. We evaluate the performance of RVH with extensive experiments on a production operated cloud platform using real rulesets and synthetic rulesets. Our results show that the packet classification performance of RVH is $15.7\times$, $14.1\times$, and $1.6\times$ on average that of TSS, PS and TM respectively.
The update speed of RVH is $3.9\times$ that of PS, $1.7\times$ that of TM and $1.1\times$ that of TSS on average.
Concerning the memory footprint, RVH occupies the space only $38\%$ that of PS, $56\%$ that of TSS and nearly the same as TM on average.
In general, the performance of RVH outperforms all existing similar hash-based packet classification algorithms.

The remainder of this paper is organized as follows. In Section~\ref{sect:relat}, we survey related work and summarize their differences from RVH. We describe the range-vector space and the algorithms in Section~\ref{sect:rvh}. We present and verify the performance model of packet classification for RVH in Section~\ref{sect:model}. In Section~\ref{sect:part}, we describe the distribution of the prefix length on each field and propose the partition policy for range-vectors. We evaluate the performance of RVH in Section~\ref{sect:exper}, and discuss the ways to further improve RVH's performance in Section~\ref{sect:discu}. Finally, our work is summarized in Section~\ref{sect:conclu}.

%% file: sect-relat.tex

\section{Related Work}
\label{sect:relat}

Existing packet classification schemes fall into three categories: the hardware based, the dimensionality reduction based, and the space partitioning based.

\textbf{Hardware Based:} \textit{Ternary Content Addressable Memory (T-CAM)}~\cite{lakshminarayanan2005algorithms, Luo2014A, SAX2016, He2017Partial} is the de facto standard chip in practice for high-speed packet classification. \textit{T-CAM} utilizes a native hardware parallel search to achieve low deterministic lookup time. However, \textit{T-CAM} suffers from limited memory size, slow rule update and power-hungry. 
Other hardware platforms proposed for packet classification include \textit{Graphics Processing Unit (GPU)}~\cite{varvello2016multilayer} and \textit{Field Programmable Gate Array (FPGA)}~\cite{jiang2012scalable,Bojie2016ClickNP}. And hardware instruction~\cite{Asai2015Poptrie}, specified chip and programming language~\cite{Bosshart2014P4} have also been proposed for packet classification. Their memory size, flexibility and cost are the barriers that prevent these solutions from widespread usage.

\textbf{Dimensionality Reduction Based:} \textit{Cross-producting}~\cite{wang2009scalable} and \textit{RFC}~\cite{gupta1999packet} split multi-dimensional rules into several single-dimensional ones to first match each individually, and then merge the temporary results of individual match. The update is slow due to each single-dimensional table should be updated for one rule update. Furthermore, the final merging process will become the performance bottleneck when the ruleset is large.

\textbf{Space Partitioning Based:} In these approaches, the whole ruleset is divided into several sub-spaces, with a small set of rules fallen into each sub-space. Instead of matching an incoming packet against the overall ruleset, the classification procedure is divided into two steps, determining the sub-space to search and matching the packet against the small sub-ruleset in the corresponding sub-space. This type of approaches further falls into two main subcategories: the decision tree based and the hash based approaches. 

The key idea of decision tree based approaches such as \textit{HiCuts (HiC)}~\cite{gupta1999hierarchical} and \textit{HyperCuts (HyC)}~\cite{singh2003packet} is to partition the search space recursively into several regions until the number of rules in each region is below a certain threshold. Due to the efficiency of the decision tree, these approaches can achieve high-speed packet classification. However, one of the shortcomings is large memory consumption because of rule replication, as some rules may need to be copied into multiple partitions. Slow and complicated rule updating is another drawback of these approaches. While \textit{EffiCuts (EC)}~\cite{vamanan2010efficuts} and \textit{SmartSplit (SS)}~\cite{He2014Meta} adopt the different policies of rule space partition to reduce the rule replication and the number of memory access based on rule distribution, they still fail to support fast update.

Existing hash based approaches, on the other hand, can achieve fast rule update, but not high-speed classification. 

\textit{Tuple Space Search (TSS)}~\cite{srinivasan1999packet}, which is implemented in OVS, partitions rules into different subsets based on tuples. A tuple is formed with a list of numbers, which represent the prefix lengths corresponding to the fields in the ruleset. 
Each subset is represented as a hash table to accelerate rule update, and packet classification is achieved with linear memory consumption. In order to classify a packet, there is a need to search all the hash tables, and the classification time will increase linearly with the number of hash tables. This will impair the classification performance seriously when dealing with large rulesets.

\textit{Pruned Tuple Space Search (PTSS)}~\cite{srinivasan1999packet} improves the performance by filtering out a number of unmatched tuples using tries. Although the number of tuples can be indeed reduced in \textit{PTSS}, merging the results is time consuming and the update operation is still complicated.

\textit{PartitionSort (PS)}~\cite{yingchareonthawornchai2016sorted} combines the benefits of both TSS and decision trees. Rather than partitioning rules based on tuples, PS partitions rules into sortable rulesets and stored them through balanced search trees. As a result, \textit{PS} can classify packet faster than TSS with the reduction of hash calculations, but at the cost of more time to process sortable ruleset. In other words, compared with TSS, \textit{PS} is faster at classifying packets at the cost of slowing rule updating. 

\textit{TupleMerge (TM)}~\cite{Daly2017TupleMerge} improves the classification of TSS by reducing the number of resulting tables. \textit{TM} defines compatibility between rules and tuples so that the rules with similar but not identical tuples can be placed in the same table. However, this approach may cause table overlaps, As a result, a rule may not be mapped to a definite table, which significantly compromises the performance of rule update. Moreover, its number of hash tables will increase over time, which also severely degrade its classification performance. For the performance improvement, all hash tables have to be reconstructed when the number of tables exceeds a certain threshold. These make it unsuitable for packet classification online. 

\textbf{Advantages of RVH:} The goal of our proposed RVH is to simultaneously support high-speed packet classification and fast rule update to meet the urgent requirement of emerging applications such as SDN and cloud computing. 
RVH builds on the key observation that the number of distinct combinations of field prefix lengths is not evenly distributed. Following this observation, RVH introduces the concept of \textit{range-vector} to balance the cost of the key hashing and the match verification upon the existence of hash conflict. No matter how the rules are updated, the number of hash tables will not increase beyond the original partition. Therefore, RVH can achieve high-speed classification and fast update at the same time.

%% file: sect-rvss.tex

\section{Range-vector Hash}
\label{sect:rvh}

Our goal in this work is to propose a new system model that can concurrently support high-performanec packet classification and fast rule update to meet the emerging need of flexibly reconfiguring network functions for value-added services.

In this section, we first introduce the concept of range-vector along with the procedures for the construction of hash tables based on range-vectors, then describe the processes of packet classification and rule update based on RVH, and finally compare the time and space complexity between RVH and other three reference schemes.

\subsection{Range-vector Hash Table}
\label{sect:rvh:cons}

\begin{table}[!t]
  \centering
  \caption{A Sample Classifier}
  \label{tab:samclassifier}
  \begin{tabular}{ccccc}
    \toprule
    Rule \# & SA & DA & Pri & Act \\
    \midrule
    0 & $100*$  & $11010$ & 2 & Fwd 0 \\
    1 & $101*$  & $1001*$ & 2 & Fwd 1 \\
    2 & $11111$ & $10000$ & 4 & Drop \\
    3 & $111*$  & $1000*$ & 2 & Fwd 4 \\
    4 & $0100*$ & $0110*$ & 3 & Fwd 0 \\
    5 & $001*$  & $01001$ & 3 & Fwd 2 \\
    6 & $00*$   & $01001$ & 3 & Drop \\
    7 & $01110$ & $*$     & 2 & Drop \\
    8 & $110*$  & $1*$    & 1 & Fwd 1 \\
    9 & $*$     & $*$     & 0 & Fwd 3 \\
    \bottomrule
  \end{tabular}
\end{table}

\begin{table}[!t]
  \centering
  \caption{Range-vectors for the Sample Classifier}
  \label{tab:samrangevector}
  \begin{tabular}{ccc}
    \toprule
    Hash Table \# & Range-vector & Rules Mapped To \\
    \midrule
    0 & $([3, 6), [4, 6))$ & 0, 1, 2, 3, 4, 5 \\
    1 & $([3, 6), [0, 4))$ & 7, 8             \\
    2 & $([0, 3), [4, 6))$ & 6                \\
    3 & $([0, 3), [0, 4))$ & 9                \\
    \bottomrule
  \end{tabular}
\end{table}

We introduce the concept of Range-vector Hash Table using an example classifier shown in Table~\ref{tab:samclassifier}. The classifier contains ten rules, each is formed with four fields: \textit{source address (SA)} and \textit{destination address (DA)} each with the maximum length $5$ bits, a \textit{priority (Pri)} field, and an \textit{action (Act)} field. For packet classification, we take the numeric fields as inputs to find the actions to take. 
In this example, for each incoming packet, we use SA and DA to search for the matching rule. 
We first introduce \textit{length-vector}. Each rule can be mapped to a length-vector, which is a vector with each element representing the length of the corresponding field in the rule. As rule \#0 has a 3-bit SA and a 5-bit DA, its length-vector is $(3, 5)$. The length-vectors of all rules in this example classifier can be summarized with a \textit{range-vector} $([0, 6), [0, 6))$. A range-vector is a vector with each element representing a length range of the corresponding field in the ruleset.
To facilitate fast rule matching, a range-vector can be further divided into several disjoint finer-grained range-vectors, each containing a set of length-vectors. For instance, the range-vector $([0, 6), [0, 6))$ can be divided into four disjoint fine-grained range-vectors $([0, 3), [0, 4))$, $([0, 3), [4, 6))$, $([3, 6), [0, 4))$ and $([3, 6), [4, 6))$. The length-vector of rule \#0, $(3, 5)$, will be mapped to the range-vector $([3, 6), [4, 6))$. Obviously, the first length $3$ falls into the range $[3, 6)$, and the second length $5$ falls into the range $[4, 6)$. In order to improve the speed of classification, in this work we divide the whole ruleset into several subsets based on the range-vectors. We apply a hash table to quickly index rules mapped to each range-vector. The rules in this example can be indexed through four hash tables based on four range-vectors in Table~\ref{tab:samrangevector}, where each table is associated with a range-vector.

\begin{figure}[!t]
\centering
 \subfigure[Length-vectors]{
   \includegraphics[width=0.42\linewidth]{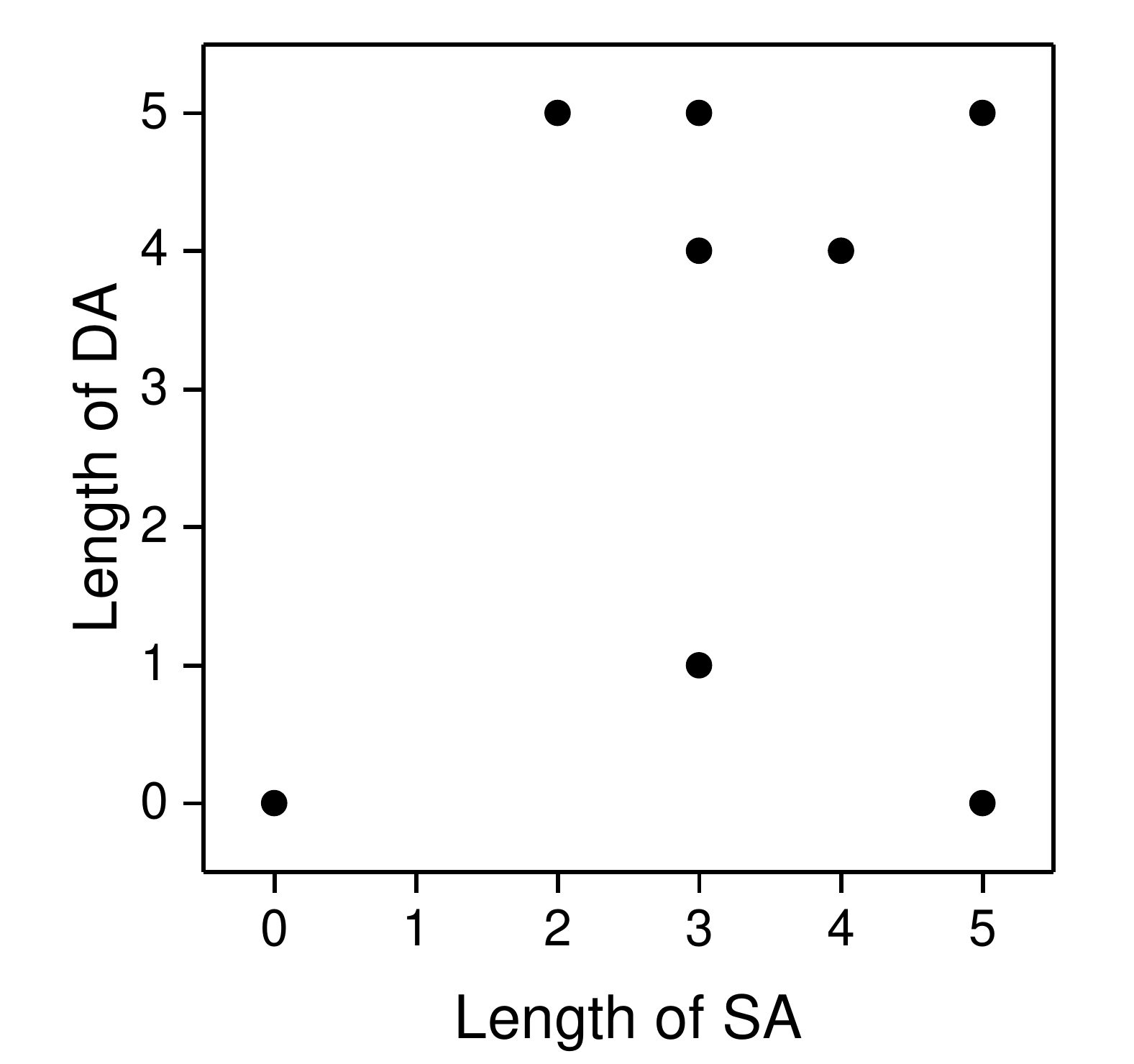}
   \label{fig:samdist:lv}
 }
 \subfigure[Range-vectors]{
   \includegraphics[width=0.42\linewidth]{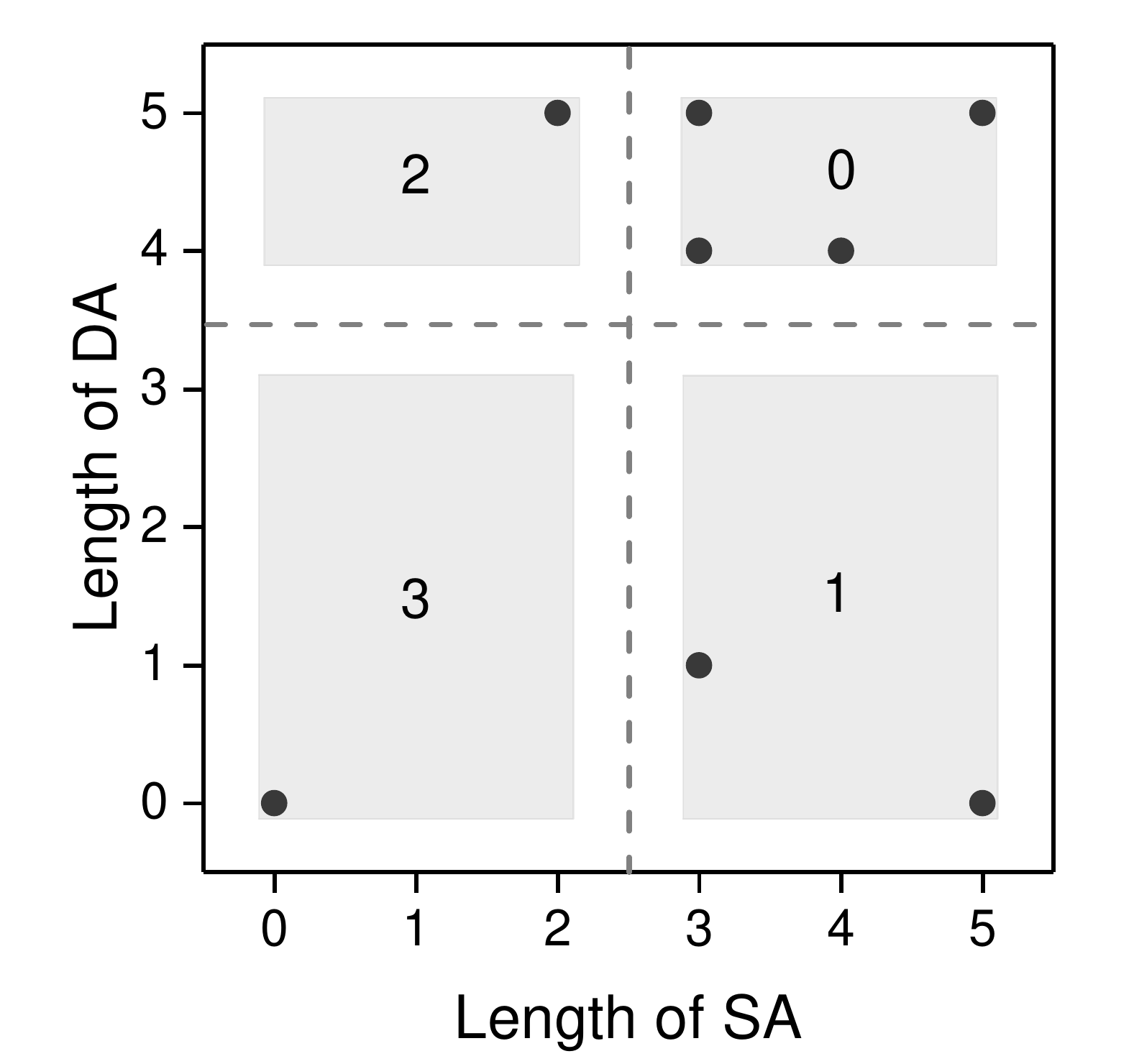}
   \label{fig:samdist:rv}
 }
 \caption{Distribution of the length-vectors (represented as points) and the range-vectors (represented as grey rectangles) of the sample classifier over the 2-dimensional space.}
 \label{fig:samdist}
\end{figure}

In Figure~\ref{fig:samdist:lv}, we plot the length-vectors corresponding to rules in Table~\ref{tab:samclassifier} over two dimensions, where each point represents a length-vector. Each range-vector is represented with a grey rectangle in Figure~\ref{fig:samdist:rv} and corresponds to a hash table, which stores a set of rules. In order to look for a rule with a hash function, we need a hash key. A natural way is to construct the key with the concatenation of field bits used for the rule matching. However, different rules may correspond to different lengths of field prefixes (e.g., SA and DA in Table~\ref{tab:samclassifier}), while we need hash keys to have the same length. To address this issue, we further introduce the concept of \textit{base-vector} to regulate the lengths of field prefixes used in the concatenation as the hash keys. More specifically, for each range element in the range-vector, we select its lower bound as a corresponding element of the base-vector to restrict the length for concatnation. As each hash table is associated with one range-vector, it is also associated with one base-vector.

\begin{table}[!t]
  \centering
  \caption{A Sample for Hash Table \#0}
  \label{tab:samtab}
  \begin{tabular}{ccc}
    \toprule
    Hash Value & Hash Key & Rules \\
    \midrule
    0x12 & $1011001$ & 1    \\
    0x34 & $0100110$ & 4    \\
    0x56 & $0010100$ & 5    \\
    0x78 & $1111000$ & 2, 3 \\
    0x90 & $1001101$ & 0    \\
    \bottomrule
  \end{tabular}
\end{table}

Table~\ref{tab:samtab} shows a sample for hash table \#0 (ignoring the empty entries) built according to the range-vector $([3, 6), [4, 6))$, and the rules \#0 $\sim$ \#5 are mapped to the hash table \#0.
The sample hash table \#0 has the base-vector $(3, 4)$, based on which the rule \#0 forms its hash key ``$1001101$'' by concatenating the first $3$ bits and the first $4$ bits of its SA (``$100*$'') and DA (``$11010$'') prefixes. The key is hashed to the example address with the offset ``0x90''. Other rules are stored at different hashed addresses using the same method.

As the use of the base-vector omits some bit information and provides a wild card for matching, two different rules may correspond to the same hash key. For example, in Table~\ref{tab:samtab}, both rules \#2 and \#3 have the same hash key ``$1111000$''. We call this phenomenon the \textit{rule overlap}. When searching the range-vector hash table, rule overlapping requires a further verification to find the exact matching rule.

\subsection{Packet Classification}
\label{sect:rvh:class}

When receiving a packet, the classifier needs to search each hash table to find the best matching rule. A packet may be matched with several rules, and we apply the priority to determine the final rule for action. Initially we set a dummy rule with the priority of 0. As a basic scheme, all existing range-vector hash tables are sequentially searched, and the matched rule with the highest priority is recorded. After searching through all tables, the classification module either returns the matched rule with the highest priority or reports that no rules are matched.

\begin{algorithm}[!t]
  \caption{Packet Classification using RVH.}
  \label{alg:class}
  \begin{algorithmic}[1]
    \Require $p$, a data packet;
    \Require $L$, a list of hash tables with priority sorting;
    \Ensure $ret$, the highest-priority matched rule;

    \State $ret \leftarrow \NULL$;
    \For{$t$ in $L$} \comm{$t$, current hash table}
      \If{$pri(ret) \geqslant pri(t)$}
        \State break;
      \EndIf
      \For{$r$ in $t[hash(p)]$} \comm{$t$, current rule}
        \If{$r$ matches $p$}
          \If{$pri(r) > pri(ret)$}
            \State $ret \leftarrow r$;
          \EndIf
        \EndIf
      \EndFor
    \EndFor
    \Return $ret$;
  \end{algorithmic}
\end{algorithm}

Although packet classification is very efficient by using the hash function, the existence of multiple rules with different priority values and the rule overlapping can compromise the performance. To further improve classification efficiency, we propose the exploration of priority sorting:
\begin{enumerate}
  \item \textbf{Hash table priority sorting.} We set the priority of a hash table as the highest priority of the rules contained in the table, and sort hash tables from high to low according to their priority. The search of range-vector hash tables can terminate once we find a matched rule whose priority is not less than the priority of the next hash table. When two hash tables have the same priority, we sort them according to their base-vector's modulus from large to small. Hash tables with larger modulus may have higher average priority, because whose rules have longer field lengths. 
  \item \textbf{Overlapped rule priority sorting.} A hash value may correspond to multiple rules due to hash collision or rule overlapping. To reduce the time needed to further determine which rule is matched, we sort the overlapped rules from high to low according to their priority during rule inserting. The verification can stop once a matched rule is found without the need of checking all the overlapped rules.
\end{enumerate}
We list the pseudocode of packet classification using RVH in Algorithm~\ref{alg:class}.

We take the classifier in Table~\ref{tab:samrangevector} as an example to classify a packet whose SA and DA fields are $(11111, 10000)$. With the hash table priority sorting, the order of the hash table list is: hash table \#0 with the priority of 4, hash table \#2 with the priority of 3, hash table \#1 with the priority of 2, and finally hash table \#3 with the priority of 0. So we search for the matching rule from hash table \#0. With the base-vector $(3,4)$, we form the hash key ``$1111000$'' and apply it to check the hash table \#0. The hashing returns two rules, \#2 and \#3. Since rule \#2 has a higher priority (as shown in Table~\ref{tab:samclassifier}), we verify it first. We find that rule \#2 matches the packet and its priority is larger than that of the next hash table. So, the search operation terminates, and rule \#2 is returned as the matched one.

\subsection{Rule Update}
\label{sect:rvh:update}

\begin{algorithm}[!t]
  \caption{Rule Insertion with RVH.}
  \label{alg:ins}
  \begin{algorithmic}[1]
    \Require $u$, a rule to insert;
    \Require $L$, a list of hash tables with priority sorting;
    \Ensure $ret$, a boolean indicating the result of insertion;

    \State $ret \leftarrow 0$;
    \State $t \leftarrow L[map(u)]$; \comm{$t$, current hash table}
    \If{$t[hash(u)] \leftarrow t[hash(u)] + u$}
      \State $pri\_sort(t[hash(u)])$
      \State $pri\_update(t)$
      \State $pri\_sort(L)$
      \State $ret \leftarrow 1$;
    \EndIf
    \Return $ret$;
  \end{algorithmic}
\end{algorithm}

\begin{algorithm}[!t]
  \caption{Rule Deletion with RVH.}
  \label{alg:del}
  \begin{algorithmic}[1]
    \Require $u$, a rule to delete;
    \Require $L$, a list of hash tables with priority sorting;
    \Ensure $ret$, a boolean indicating the result of deletion;

    \State $ret \leftarrow 0$;
    \State $t \leftarrow L[map(u)]$; \comm{$t$, current hash table}
    \If{$t[hash(u)] \leftarrow t[hash(u)] - u$}
      \State $pri\_update(t)$
      \State $pri\_sort(L)$
      \State $ret \leftarrow 1$;
    \EndIf
    \Return $ret$;
  \end{algorithmic}
\end{algorithm}

Given an existing classifier, we now describe the rule update operation (insertion and deletion) in RVH. Inserting a rule into the classifier is rather simple. We first identify which range-vector hash table the rule should be inserted into, according to the prefix lengths specified by the base-vector. We then insert the rule into the correct location by hashing its key. If there are other rules at this location, these overlapped rules will be reordered according to their priority. The pseudocode for the rule update operation, including rule insertion and deletion, is shown in Algorithms~\ref{alg:ins} and \ref{alg:del}.

For instance, suppose we insert rule \#10 into the classifier in Table~\ref{tab:samrangevector}. The fields of rule \#10 are $(011*, 011*)$ and its priority is 3. It should be inserted into the hash table \#1 and overlaps with rule \#7 with the same hash key ``$011$''. Because rule \#10 has a higher priority, it is inserted before rule \#7.

Deletion of a rule from the classifier is even simpler: locating the rule in the corresponding hash table, and then deleting it. If the corresponding hash table is empty after the deletion, we also delete the hash table. For instance, if we delete rule \#9 from the hash table \#3 as indicated by Table~\ref{tab:samrangevector}, we also delete the instance of this hash table as it is empty after the deletion of rule \#9.

\subsection{Time and Space Complexity}
\label{sect:rvh:complex}

We compare the time and space cost of RVH with other three state-of-the-art approaches, and the result is shown in Table~\ref{tab:complex}. We assume that the ruleset consists of $n$ rules with $d$ fields. The ruleset is implemented as $m$ tuples (i.e., $m$ length-vectors) in TSS, $m'$ partitions in PS, $m''$ tuples in TM, and $m'''$ range-vectors in RVH.

\begin{table}[!t]
  \centering
  \caption{Cost comparison in terms of classification time, update time and memory footprint}
  \label{tab:complex}
  \begin{tabular}{llll}
    \toprule
    Algo. & Classification & Update & Mem. \\
    \midrule
    TSS~\cite{srinivasan1999packet}            & $O(dm)$             & $O(d)$              & $O(dn)$ \\
    PS~\cite{yingchareonthawornchai2016sorted} & $O(dm' + m'\log n)$ & $O(dm' + m'\log n)$ & $O(dn)$ \\
    TM~\cite{Daly2017TupleMerge}               & $O(dm'')$           & $O(dm'')$           & $O(dn)$ \\
    RVH                                        & $O(dm''')$          & $O(d)$              & $O(dn)$ \\
    \bottomrule
  \end{tabular}
\end{table}

As range-vectors in RVH are formed with the effective grouping of length-vectors that are the tuples in TSS, $m'''$ is far less than $m$. $m$ is typically of the order of hundred in practice, while $m'''$ is often two order of magnitude less. Besides, TM often has more hash tables than that of RVH during the system running. As far as we know, RVH can achieve the highest packet classification performance, and is also the only approach that has the update performance comparable with TSS. Our experiments in Section~\ref{sect:exper} confirm that RVH has the superior performance to meet the requirement for a modern classifier.

%% file: sect-model.tex

\section{Performance Model}
\label{sect:model}

In order to optimize RVH with high classification and update performance, we need to determine the number of range-vectors thus the number of hash tables to build, and the policy to cover the whole ruleset with range-vectors. As a guide, we propose a performance model to analyze the bottlenecks and factors that impact the performance. We verify the accuracy of our model with experimental results.

\subsection{Overview of RVH}
\label{sect:model:overview}

In RVH, we have introduced the concept of range-vector. A set of disjoint range-vectors partition the matching space (i.e., the whole ruleset), with each range-vector associated with a hash table for the quick match of a packet against rules contained in that hash table.

In general, classifying a packet using RVH consists of three procedures: 1) searching all range-vector hash tables, and for each table a simple hash operation is performed to determine if there are rules matching the header of the incoming packet; 2) verifying the exact rules matched in a hash table if there exist several ones stored in the same hashed address to avoid the false positive caused by hash conflict or rule overlap; and 3) determining the rule with the highest priority among the verified ones.

When updating a rule, the hash table where the rule should be contained is determined first. Inserting a rule into a range-vector hash table takes only two steps: 1) mapping the rule to the corresponding range-vector (hash table), or creating one if the hash table does not exist; and 2) storing the rule at the proper entry of the hash table determined by the hash operation.
The process of deleting a rule from range-vector hash table is similar to the inserting procedure.

\subsection{Analytical Performance of RVH}
\label{sect:model:formula}

Based on the logic and process of RVH approach described above, we establish its performance model for packet classification to formally explore the performance bottleneck of RVH. Table~\ref{tab:samvar} lists the notations used in our analysis.

\begin{table}[!t]
  \centering
  \caption{Symbol and Definition}
  \label{tab:samvar}
  \begin{tabular}{cl}
    \toprule
     Sym. & Definition \\
    \midrule
     $h_i$ & Hash calculation time of the $i$-th hash table \\
     $c_i$ & Match verification time of the $i$-th hash table \\
     $r_i$ & Average hit ratio in the $i$-th hash table \\
     $o_i$ & Average overlap ratio of the $i$-th hash table \\
     $\bar{h}$ & Average hash calculation time\\
     $\bar{c}$ & Average comparing time for verification \\
     $\bar{q}$ & Average priority comparing time \\
     $e_i$ & The number of filled entries in the $i$-th hash table \\
     $s_i$ & Size of the $i$-th hash table \\
     $n_i$ & The number of rules in the $i$-th hash table \\
     $m$   & The number of hash tables (range-vectors) \\
    \bottomrule
  \end{tabular}
\end{table}
The total time $T$ taken to classify an incoming packet with a classifier consists of three parts: the time for hashing, for match verification and for priority selection. The time for hashing is decided by the number of hash tables (range-vectors) and the cost of each hash function. The match verification should be undertaken for each hash table matched with multiple rules to avoid the false positive from hash collision or rule overlap. The amount of match verification for an incoming packet equals to the number of matched entries. Given the specific refined hash function, the probability of matching an entry in a hash table is proportional to the utilization ratio of each hash table. The utilization ratio is defined as the number of entries divided by the size of the hash table.

The rules with the same hash value have to be further checked to eliminate the ones with false positive, and the rule with the highest priority is selected when multiple ones are matched. Therefore, the rule overlap will increase time for match verification. The average time spent for rule verification is the time cost for a rule verification weighted by the average overlap ratio, which is defined as the number of rules in the hash table divided by its the number of entries.
Consequently, the time for packet classification in RVH is
\begin{equation}
\label{eq:1}
  T = \sum_{i=1}^{m} h_i + \sum_{i=1}^{m}c_i r_i o_i + \bar{q}
\end{equation}
It is noteworthy that the time for hashing of any hash table (i.e., $h_i$) is close to that of other hash table, so does the time for verification. Taking the implementation in OVS as an example, the size of each hash table can be dynamically adjusted according to the number of entries with hash values, so the utilization ratio of each hash table is kept around the average of those of all hash tables. 
The utilization ratio and the overlap ratio can be calculated based on the number of rules, the number of entries and the size of hash table as follows:
\begin{equation}
\label{eq:2}
  r_i = \frac{e_i}{s_i}
\end{equation}
and
\begin{equation}
\label{eq:3}
  o_i = \frac{n_i}{e_i}
\end{equation}

Replacing the hit ratio and the overlap ratio in (\ref{eq:1}) with (\ref{eq:2}) and (\ref{eq:3}), the time cost to classify a packet using RVH can be converted to
\begin{equation}
\label{eq:4}
  T = \sum_{i=1}^{m} h_i + \sum_{i=1}^{m}c_i \cdot \frac{e_i}{s_i} \cdot \frac{n_i}{e_i} + \bar{q}
\end{equation}
which can be simplified to
\begin{equation}
\label{eq:5}
  T = m \bar{h} + \bar{c} \sum_{i=1}^{m}\frac{n_i}{s_i} + \bar{q}.
\end{equation}

The number of entries in the $i$-th hash table ($e_i$) and the size of the $i$-th hash table ($s_i$) are related to the design of the hash. Besides, the number of hash tables ($m$) depends on the distribution of the prefix length in each dimension. Given a specific hardware platform, the average time for hashing $\bar{h}$, for verification $\bar{c}$ and for priority comparing $\bar{q}$ are often invariant.
Therefore, based on Eq.~\ref{eq:5}, we can conclude safely that both the number of hash tables and the overlap ratio in individual tables determine the classification performance of RVH: the fewer the number of hash tables produced by the ruleset and the lower overlap ratio, the higher the performance.

\subsection{ Verification of the Performance Model}
\label{sect:model:veri}

We verify the above performance analysis via experiments. To this end, we use the RVH core module implemented in OVS without any flow cache. The detailed experiment setup and the rulesets can be found in Section~\ref{sect:exper}.

\begin{table}[!t]
  \centering
  \caption{Estimated Time of Classifying a Packet}
  \label{tab:testmodel}
  \begin{tabular}{ccccc}
    \toprule
    Rulesets & $m$ & Saturation & $T$~($\mu s$) & Error~(\%) \\
    \midrule
    ACL1   & 703 & 0.71 & 45.23 & 4.31 \\
    ACL2   & 702 & 0.71 & 45.16 & 5.28 \\
    FW1    & 181 & 0.73 & 11.66 & 6.90 \\
    FW2    & 179 & 0.74 & 11.54 & 6.67 \\
    IPC1   & 71  & 0.80 & 4.59  & 11.64 \\
    IPC2   & 71  & 0.82 & 4.60  & 10.07 \\
    Cloud1 & 226 & 0.72 & 14.55 & 6.45 \\
    Cloud2 & 251 & 0.77 & 16.22 & 8.37 \\
    \bottomrule
  \end{tabular}
\end{table}
Firstly, we tested one billion times of hashing, match verification and priority comparison, and obtained the average time of hashing a packet ($\bar{h}$), the average time of verifying a match ($\bar{c}$), as well as the average time of comparing priority ($\bar{q}$), which are $61.0$ns, $4.7$ns and $0.9$ns respectively. We then determined the utilization ratio and the overlap ratio of each hash table. The average overlapping ratio equals to $(1/m) \cdot \sum_{i=1}^{m}n_i/s_i$ and is denoted as \textit{saturation} in the result Table~\ref{tab:testmodel}. Based on these parameters, we can estimate the time of classifying a packet based on Eq.~(\ref{eq:5}).

\begin{figure}[!t]
  \centering
  \includegraphics[width=0.7\linewidth]{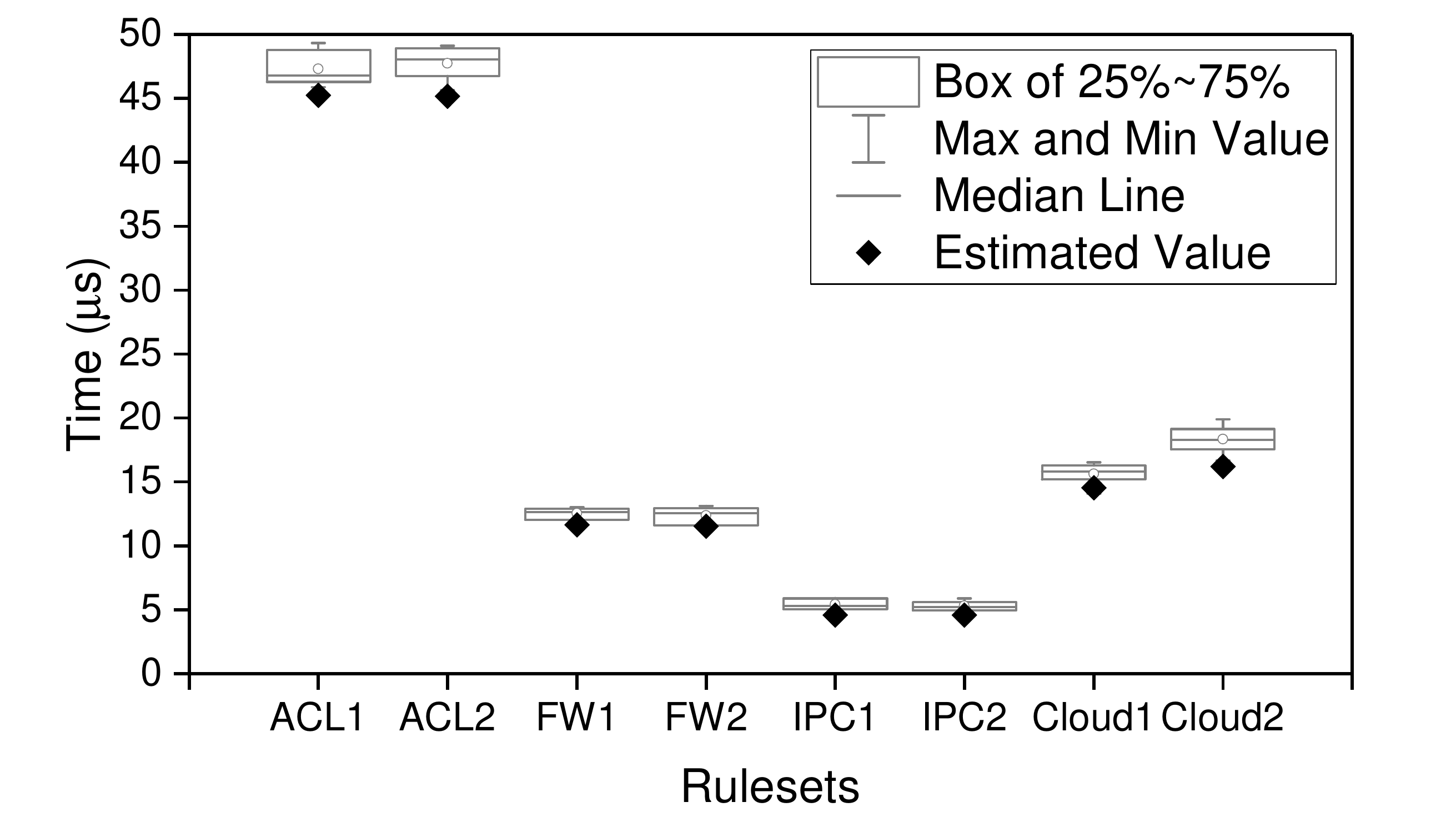}
  \caption{Comparison of the theoretical and experimental results for packet classification with RVH. The (theoretically) estimated performance is generated by our analytical model (Eq.~\ref{eq:5}), while the experimental results are obtained using the RVH implementation in OVS.}
  \label{fig:veriformula}
\end{figure}
Secondly, we run experiments with the RVH implementation in OVS without flow cache over the rulesets in Table~\ref{tab:ribs} to obtain the classification performance in practice. Figure~\ref{fig:veriformula} shows the quantile time used to classify a packet over 10 runs (the boxes and whiskers), along with the theoretically estimated time (diamond points). The average relative error between estimated time and average actual time is also listed in Table~\ref{tab:testmodel}.
We can visually see that the estimated time taken to classify a packet is very close to the actual values, though the actual value is slightly higher than the estimated one. This is because that additional calculations and CPU cache replacement (which our model does not capture) may take extra time, which is indeed not negligible.

In summary, we can safely make the following conclusions from the above analysis: 1) our performance model captures the major components of RVH, and provides reasonably precise estimation for its performance; 2) the key to significantly improve RVH's classification performance is to reduce the number of hash tables (i.e., $m$) while maintaining a low overlap ratio.

For the same reason, the main idea of both PS and TM is to reduce the number of hash tables to improve their packet classification performance. However, PS combines the decision tree, which makes the updates more difficult and results in additional overhead. While for TM, a rule may be mapped into different tables, depending on the table utilization. In addition, as time goes on, the number of tables increases, which will trigger an overall table reconstruction. These processes cause a dramatic drop in the update speed of TM. With the proper determination of the number of hash tables to use based on our performance model, RVH can achieve a high classification speed while maintaining the fast update property. We will discuss how to partition the whole ruleset into range-vectors in the next section.

%% file: sect-part.tex

\section{Range-vector Partition Policy}
\label{sect:part}

From our performance model and experimental studies in Section~\ref{sect:model}, the classification speed is highly impacted by the number of hash tables. In this section, we first introduce our key observations on the features of ruleset, and then propose the distribution-based policy to partition the packet matching space for hashing in RVH. Finally, we study the stability of prefix-length distribution in each dimension.

\subsection{Tradeoff for Partition}
\label{sect:part:tradeoff}

From our performance model, the time taken for packet classification depends on both the number of hash tables (range-vectors) and the rule overlap ratio in individual tables. These two factors are conflicting: the more the number of hash tables, the lower the chance of rule overlapping in each table. At one extreme, if we put all rules into one hash table and all rules are also mapped to the same hash entry, classifying a packet requires the search of the rule list linearly. At the other extreme, if we let each range-vector contain only one length-vector, then the classification approach is essentially TSS.

\begin{figure}[!t]
  \centering
  \subfigure[ACL]{
    \includegraphics[width=0.42\linewidth]{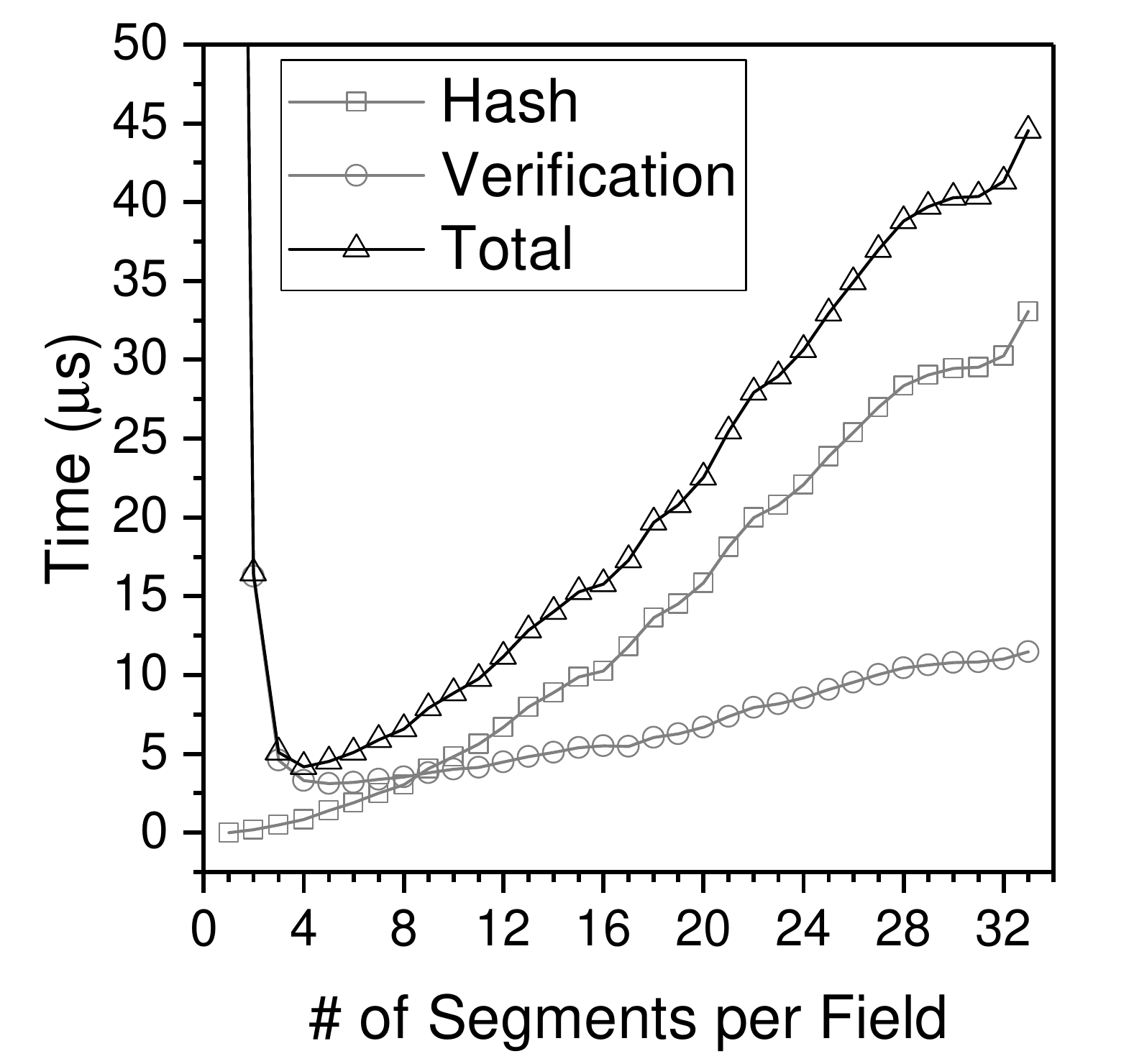}
    \label{fig:split:acl}
  }
  \subfigure[FW]{
    \includegraphics[width=0.42\linewidth]{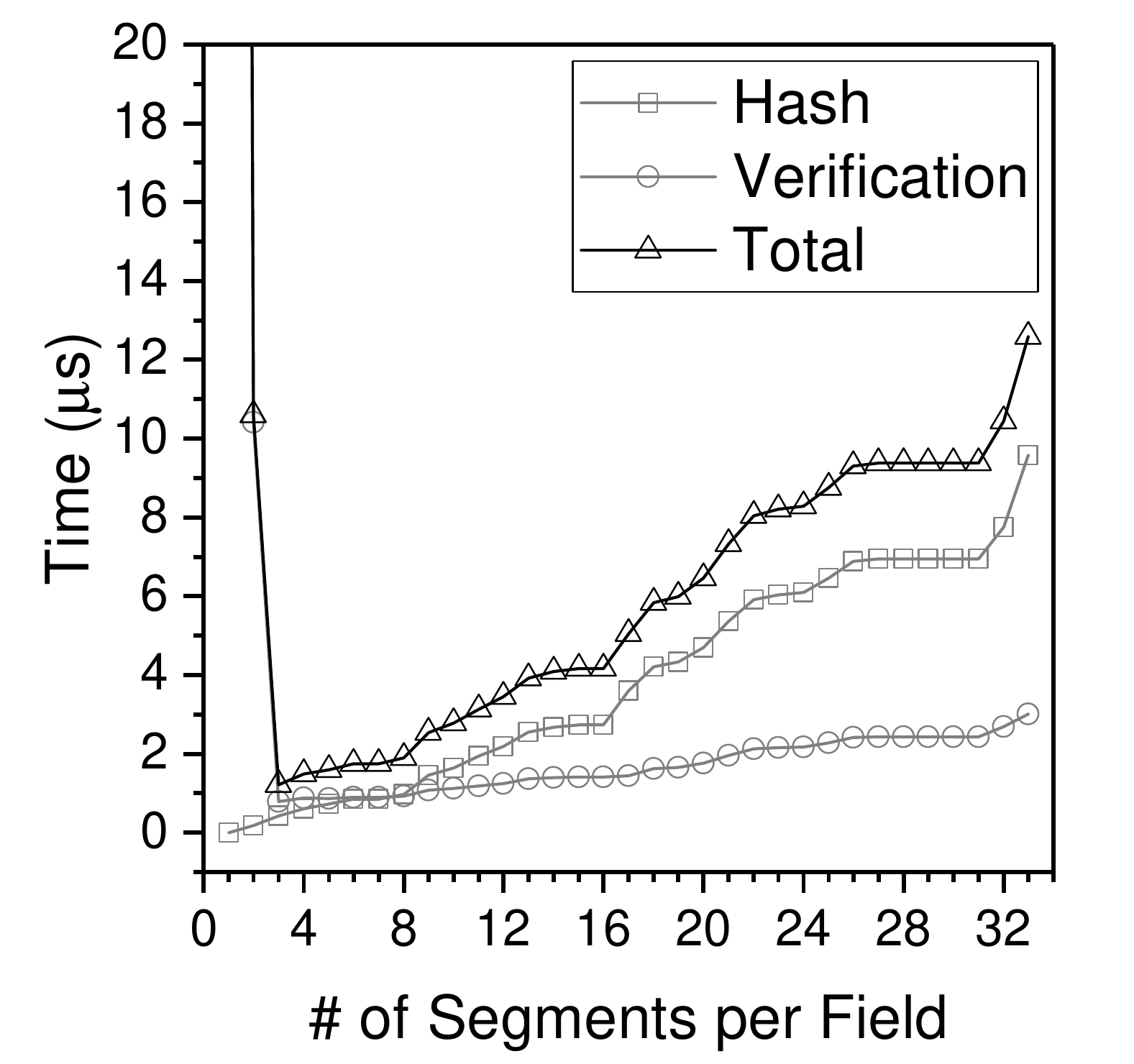}
    \label{fig:split:fw}
  }
  \subfigure[IPC]{
    \includegraphics[width=0.42\linewidth]{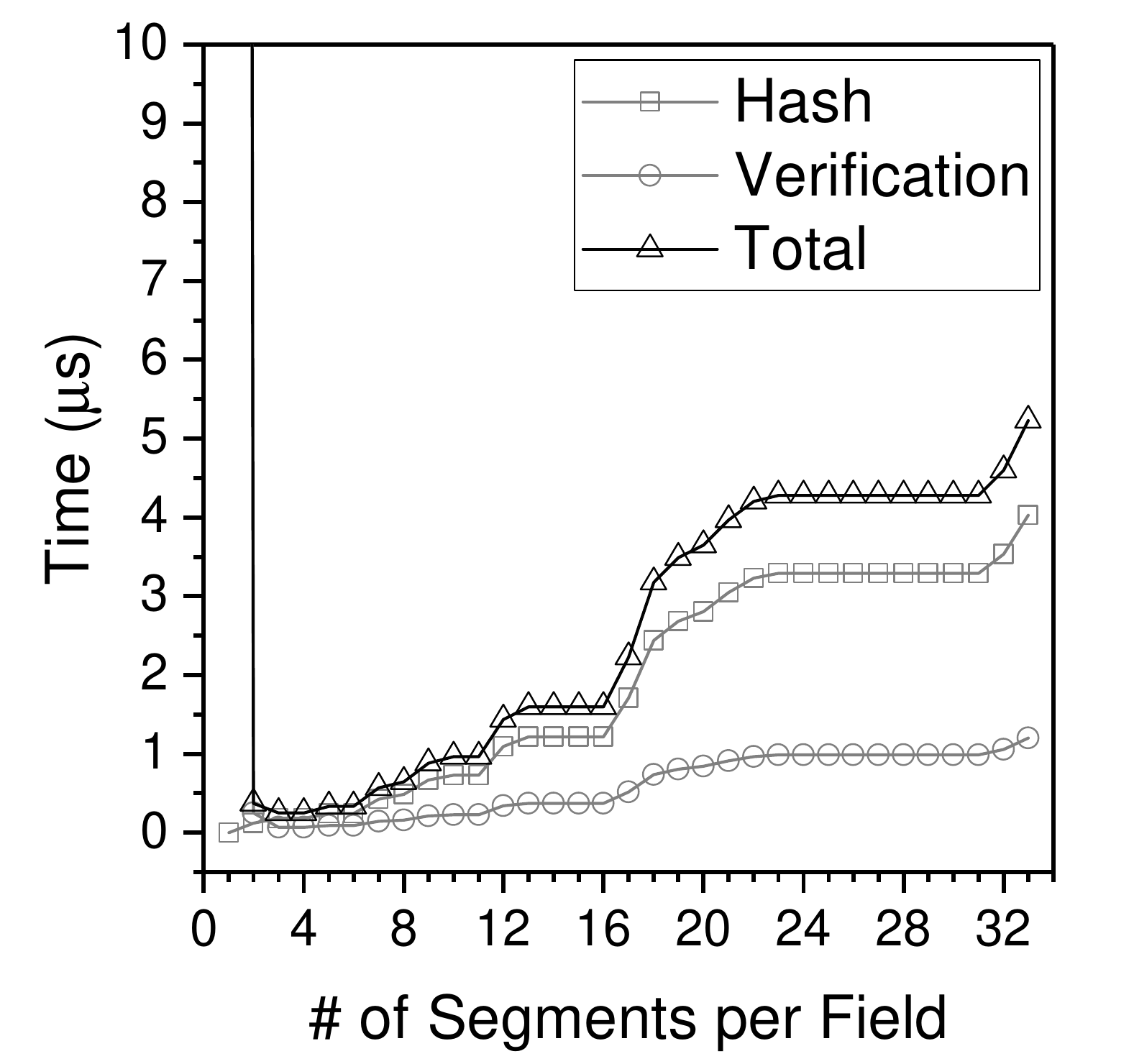}
    \label{fig:split:ipc}
  }
  \subfigure[Cloud]{
    \includegraphics[width=0.42\linewidth]{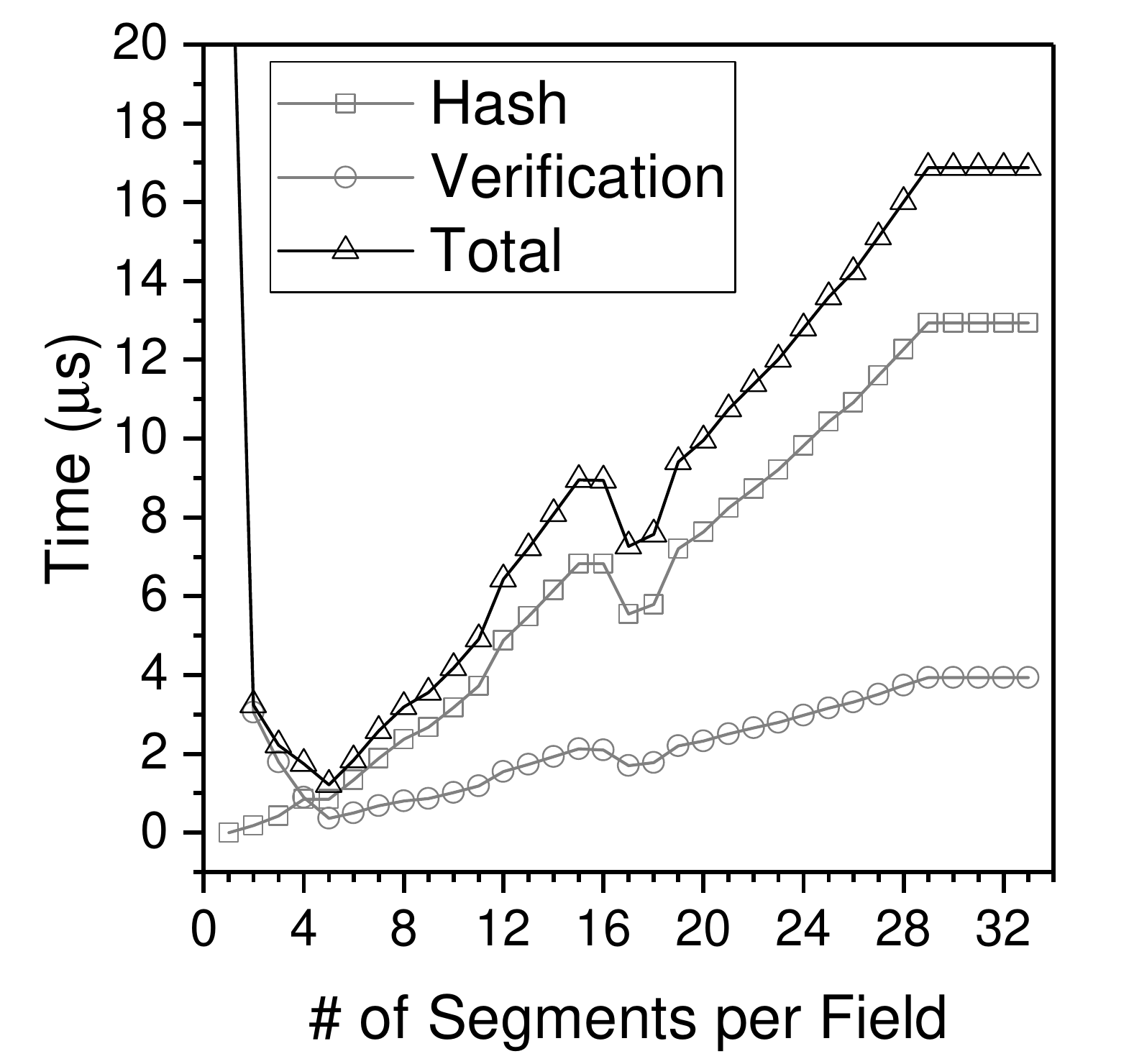}
    \label{fig:split:real}
  }
  \caption{Variation of time cost with the increase of the number of segments that each dimension (field) is split. The number of range-vectors ($m'$) is equal to $x^2$.}
  \label{fig:split}
\end{figure}

The classification time will reach the minimum if we can find the ideal balance between the two factors. In practice, it is hard to represent the relationship of the two with a close-form equation. We resort to experiments to show their interaction. We use the rulesets in Table~\ref{tab:ribs} to classify packets based on the source address (SA) and destination address (DA).
We partition the 2-dimensional overall rules evenly into a series of range-vectors, with each dimension split into a number of segments.
We report the time for hashing (i.e., $m\bar{h}$ in Eq.~\ref{eq:5}), the time for match verification (i.e., $\bar{c} \sum_{i=1}^{m}n_i/s_i$ in Eq.~\ref{eq:5}), and the total time cost (i.e., $T$ in Eq.~\ref{eq:5}) in Figure~\ref{fig:split}, where $x$-axis represents the number of segments in each dimension.
Note that $\bar{h}$, $\bar{c}$ and $\bar{q}$ are taken from Section~\ref{sect:model}, where $\bar{h}=61.0$ns, $\bar{c}=4.7$ns and $\bar{q}=0.9$ns.

As expected, for each ruleset, there is a ``sweet point'' of the number of segments to split in each dimension, where the total time cost reaches the minimum. This ``sweet point'' is ruleset dependent, but falls into the range between 3 and 5. With this range of segments in one dimension, the number of two-dimensional regions corresponding to the number of range-vectors (hash tables) varies from 9 to 25. A value smaller than the ``sweet point'' yields an extraordinarily high time cost for match verification, which dominates the total time cost. Beyond this point, the total time for verification grows slowly,  but the time for hashing increases quadratically with the number of segments.

\subsection{Distribution-based Partition}
\label{sect:part:policy}

The range-vector formed with the overall lengths of prefixes to match is partitioned into multiple range-vectors with two features: 1) all range-vectors together covers the overall lengths of prefixes to match, that any rule can be mapped to one of range-vectors; and 2) range-vectors do not intersect with each other, so that a rule can be mapped to only one range-vector.

We follow the principles below for the division of range-vectors:
\begin{enumerate}
  \item The number of range-vectors should be as small as possible. The number of hash calculations for packet classification is proportional to the number of range-vectors. Our experimental results show (see Figure~\ref{fig:split}) that keeping this number within the ranges $[9, 25]$ can achieve fast classification.
  \item The number of overlapped rules in hash tables should be kept as few as possible to reduce the time taken for verifying the match (see Figure~\ref{fig:split}). As the number of range-vectors decreases, each hash table will contain more rules, which raises the chance of rule overlap.
  \item The field lengths of rules that are mapped to a range-vector should be as close as possible to that specified by the corresponding base-vector. In other words, for each field of a rule that is used for matching, its length should be as close to the lower bound of the corresponding range as possible. Otherwise, as we use the cut-short prefix to form the hash key, it will increase the likelihood of rule overlap. 
\end{enumerate}

To find the best construction of range-vectors that meets the above principles, we may apply mathematical tools such as those based on linear programming. However, these tools often take a long time to run and are not applicable for online operation. In this paper, we propose a simple yet effective construction method.

\begin{figure}[!t]
  \centering
  \subfigure[ACL]{
    \includegraphics[width=0.42\linewidth]{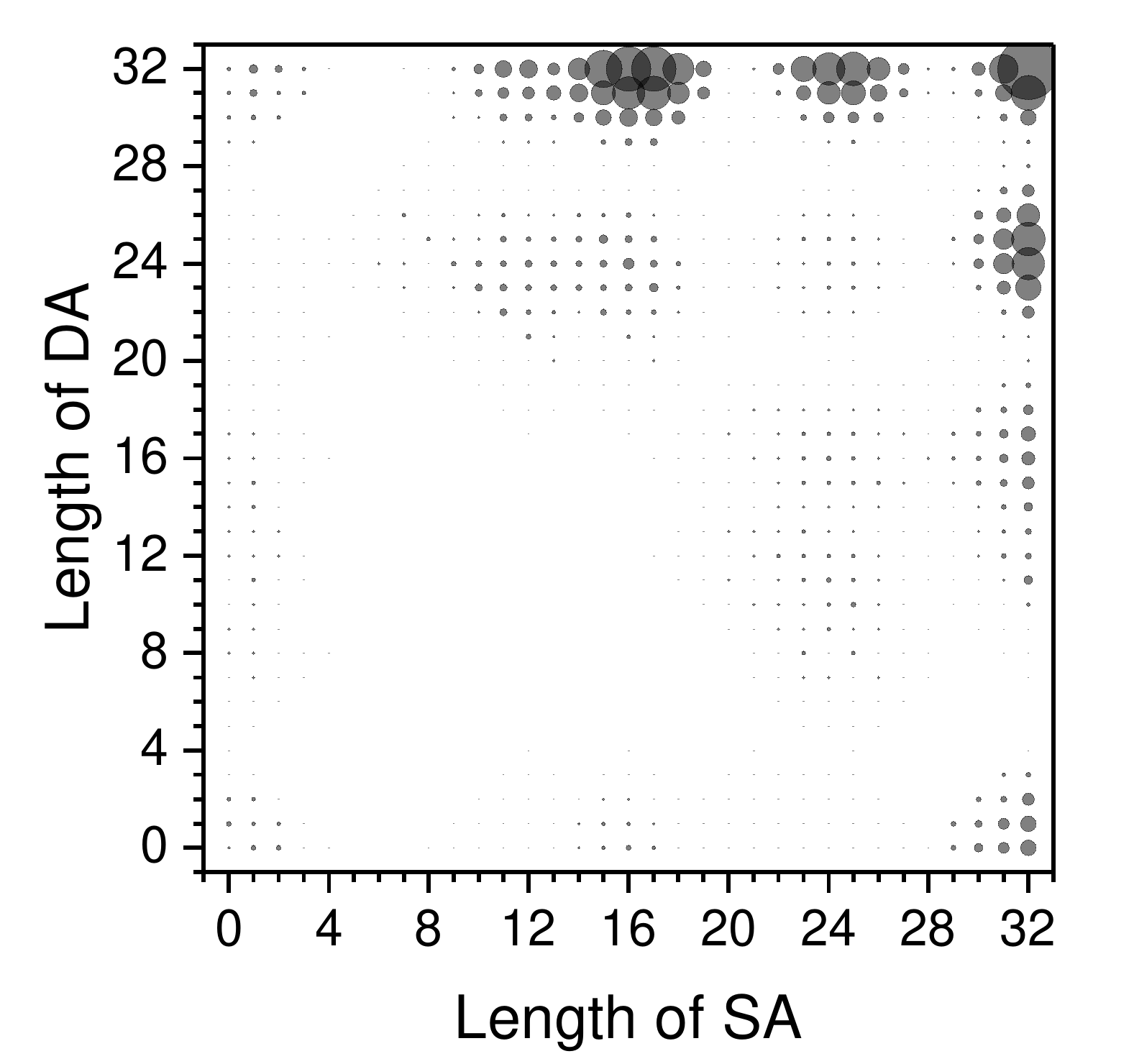}
    \label{fig:dist:acl}
  }
  \subfigure[FW]{
    \includegraphics[width=0.42\linewidth]{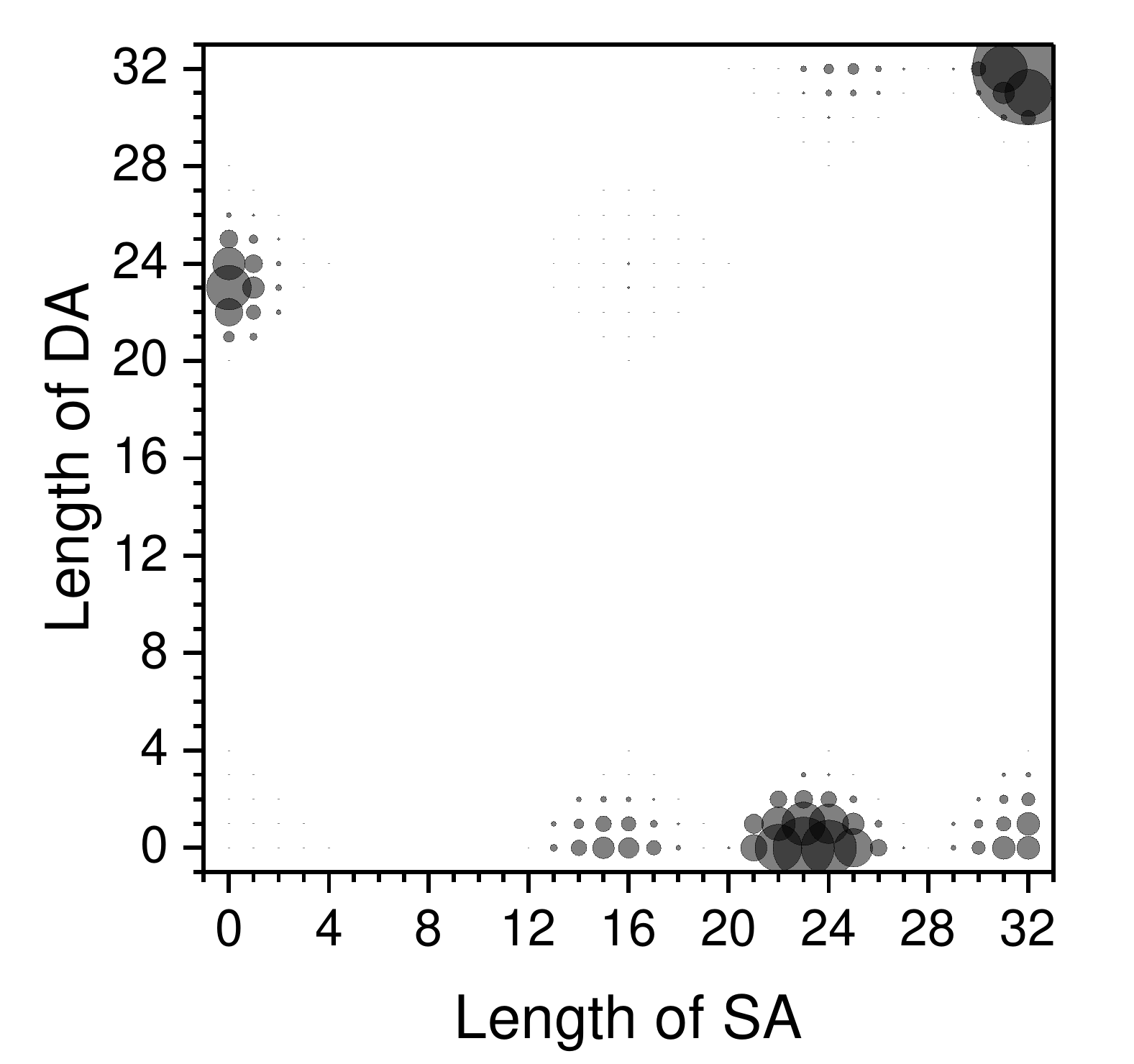}
    \label{fig:dist:fw}
  }
  \subfigure[IPC]{
    \includegraphics[width=0.42\linewidth]{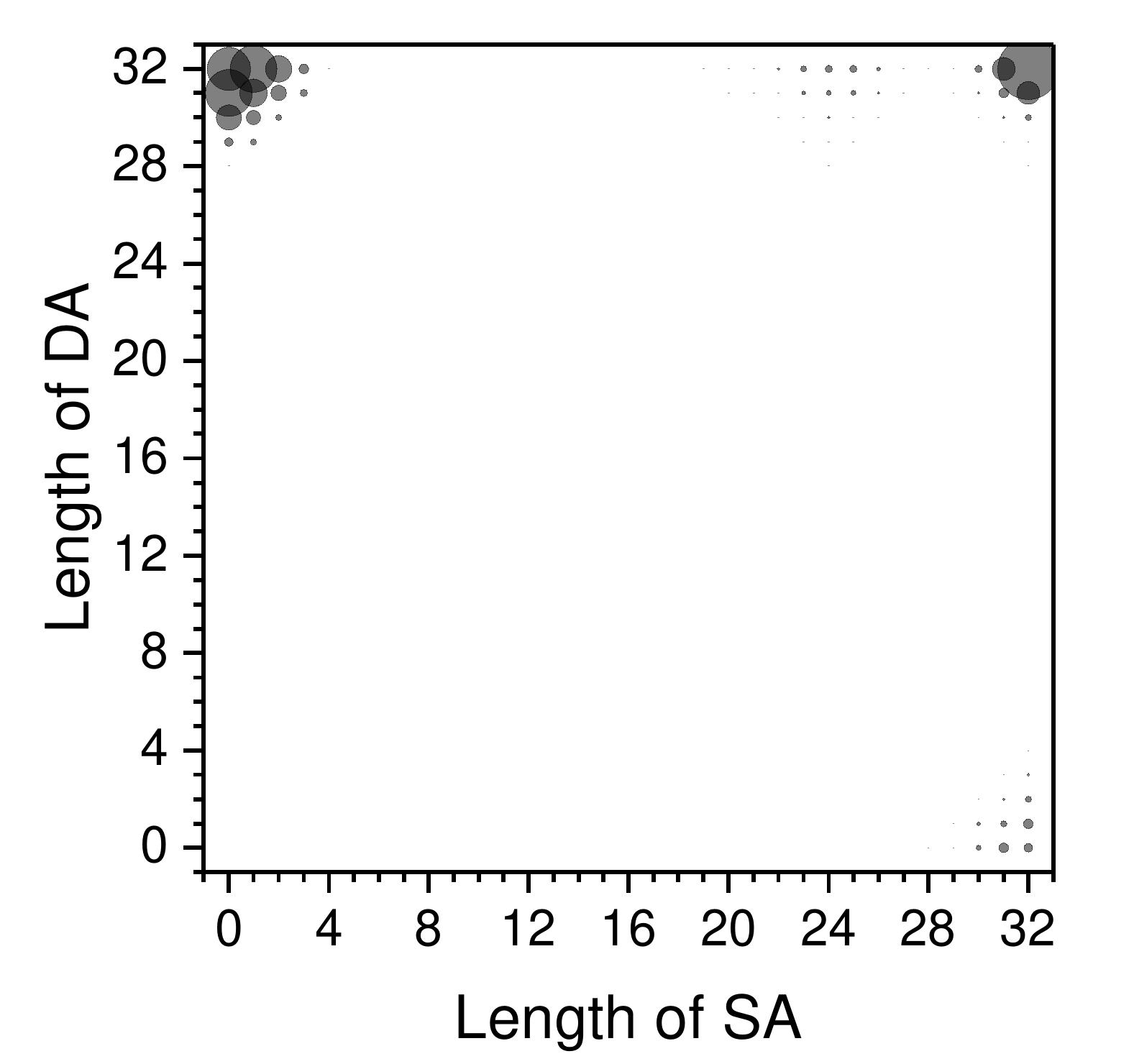}
    \label{fig:dist:ipc}
  }
  \subfigure[Cloud]{
    \includegraphics[width=0.42\linewidth]{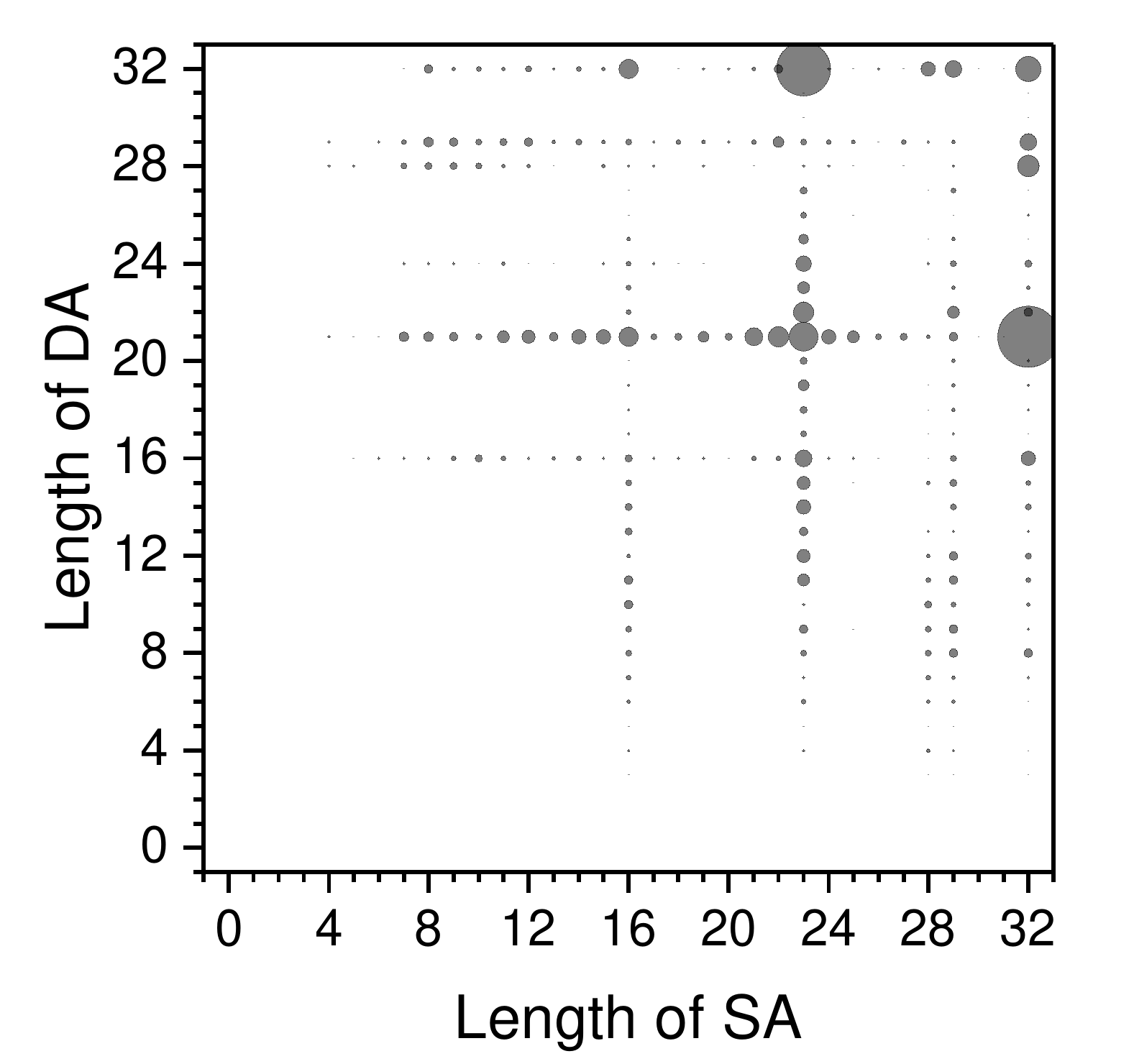}
    \label{fig:dist:real}
  }
  \caption{The distribution of the combinations of the prefix lengths of SA and DA. Each bubble represents a combination and the size of which depicts its quantity.}
  \label{fig:dist}
\end{figure}

We first examine the distribution of prefix length combinations, taking SA and DA of the rulesets in Table~\ref{tab:ribs} as examples. From the results in Figure~\ref{fig:dist}, the distribution is not uniform in the 2-dimensional space, but clustered in several ranges. This clustering behavior can be observed from all four rulesets we examined. This is the result of two facts. First, IP addresses are often assigned in blocks with several typical prefix lengths (e.g., 16-bit prefix, 24-bit prefix). Second, network functions are configured to work on IP blocks. For instance, firewall policies may block all source IP addresses from some blocks, which forms the left-most cluster in Figure~\ref{fig:dist}.
We construct range-vectors following the clustering behavior. As the number of clusters is limited, there will be only several range-vectors, which effectively reduces the number of hash tables. In addition, we can align the range-vectors to allow the rules to be close to the lower bounds of range-vectors. Finally, more rules are likely to fall into a range whose duration is large. Setting the range based on the cluster size helps to reduce the range-size thus the chance of rule overlapping.

\begin{figure}[!t]
  \centering
  \subfigure[ACL]{
    \includegraphics[width=0.42\linewidth]{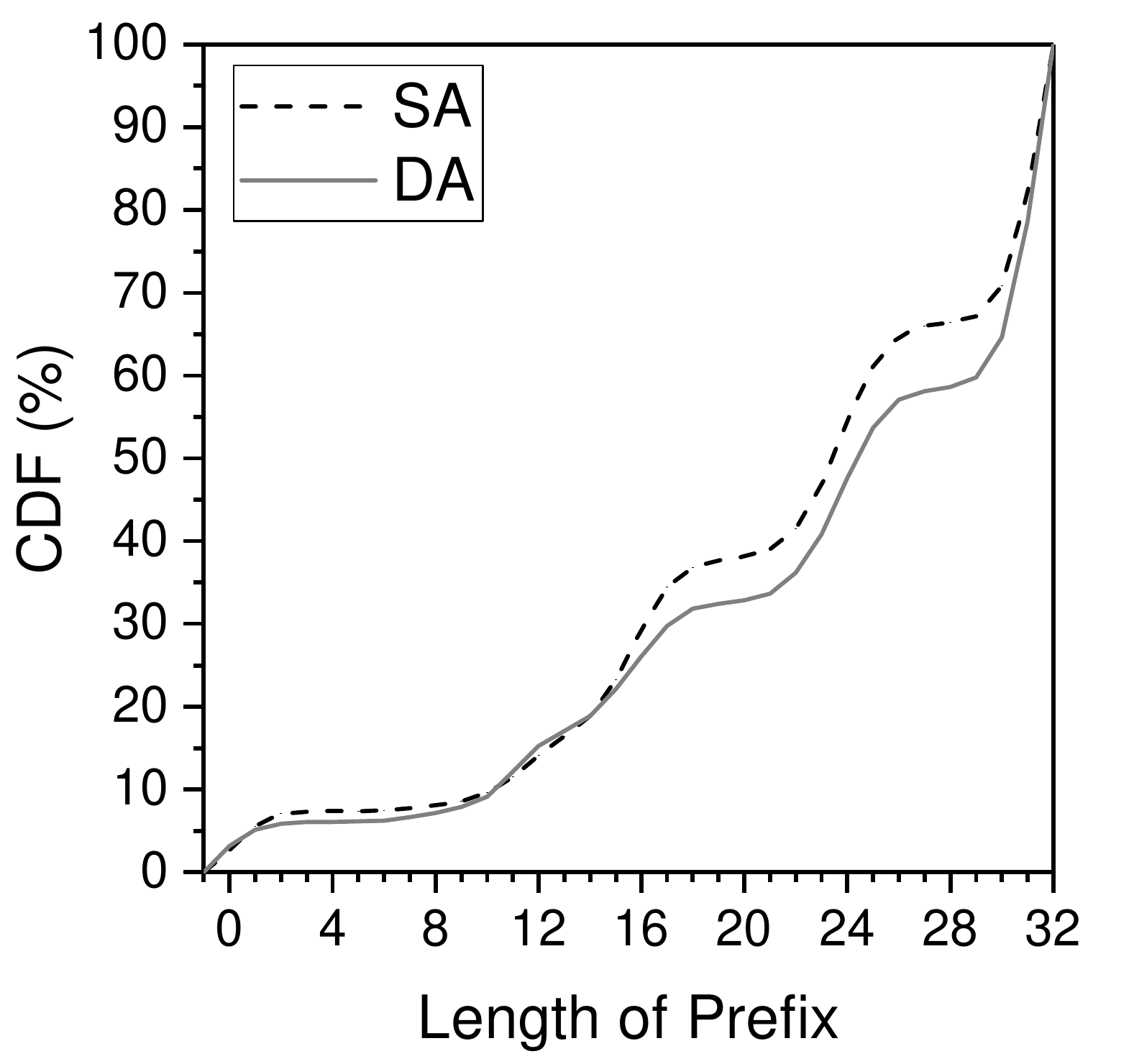}
    \label{fig:cdf:acl}
  }
  \subfigure[FW]{
    \includegraphics[width=0.42\linewidth]{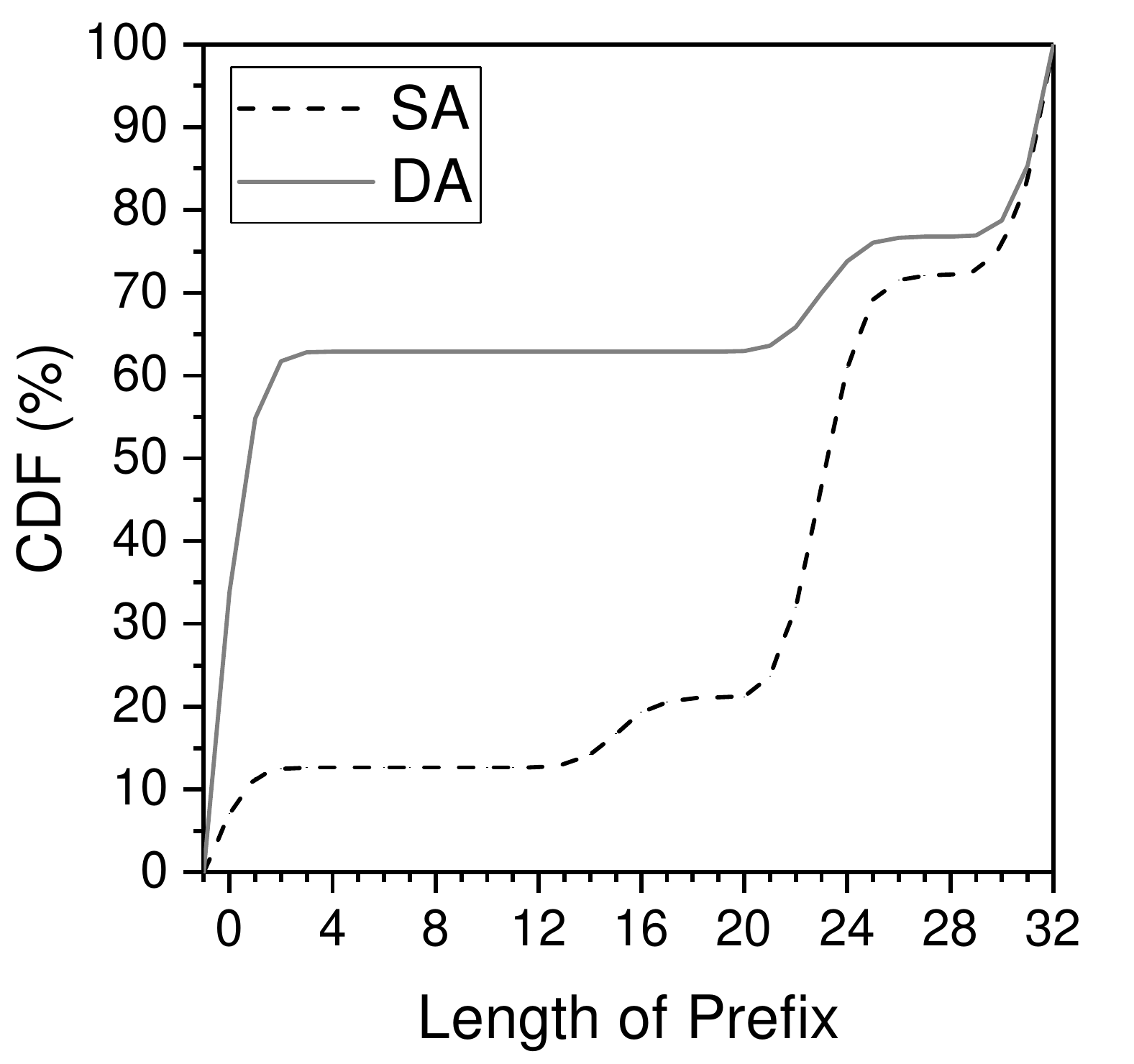}
    \label{fig:cdf:fw}
  }
  \subfigure[IPC]{
    \includegraphics[width=0.42\linewidth]{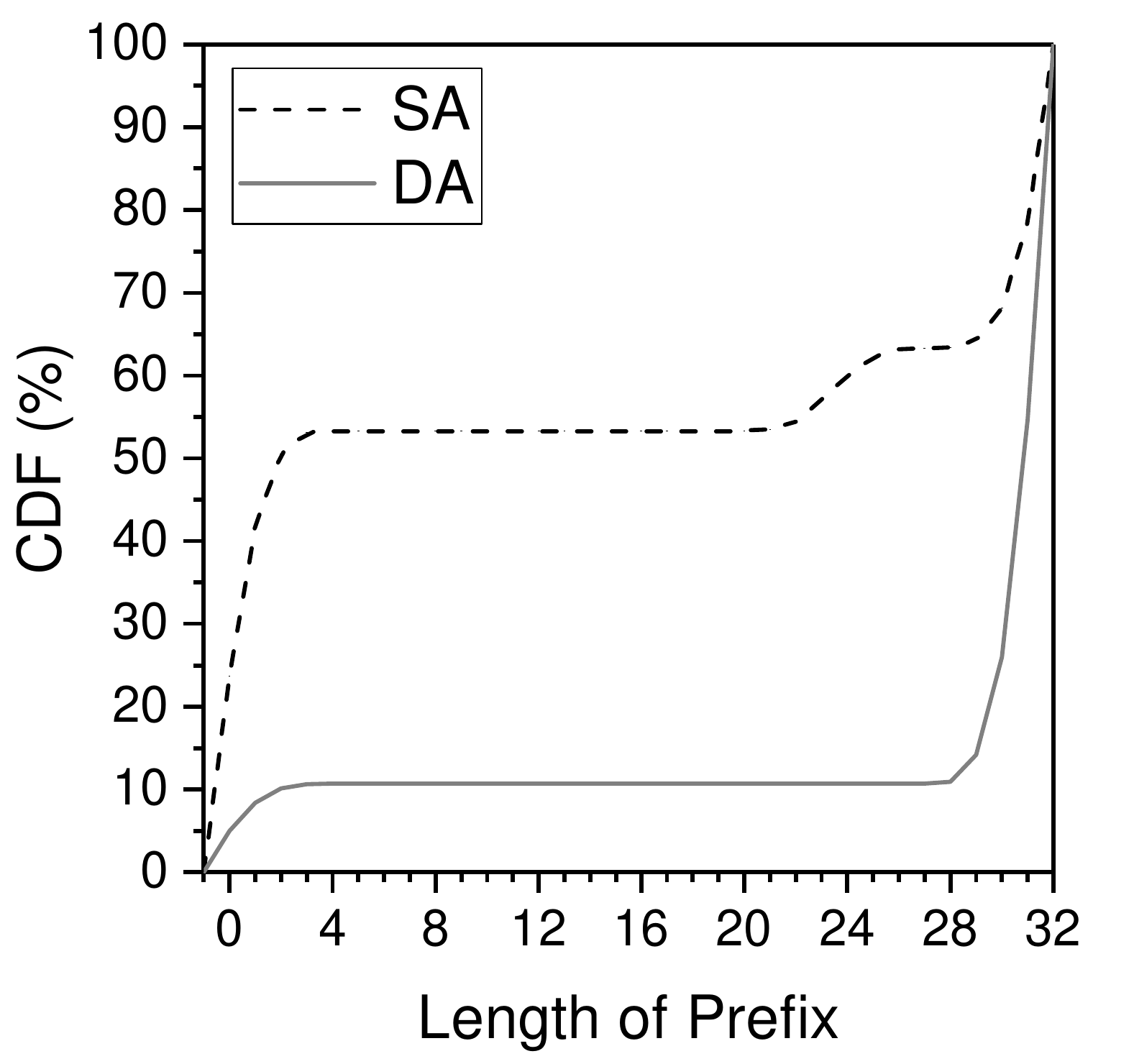}
    \label{fig:cdf:ipc}
  }
  \subfigure[Cloud]{
    \includegraphics[width=0.42\linewidth]{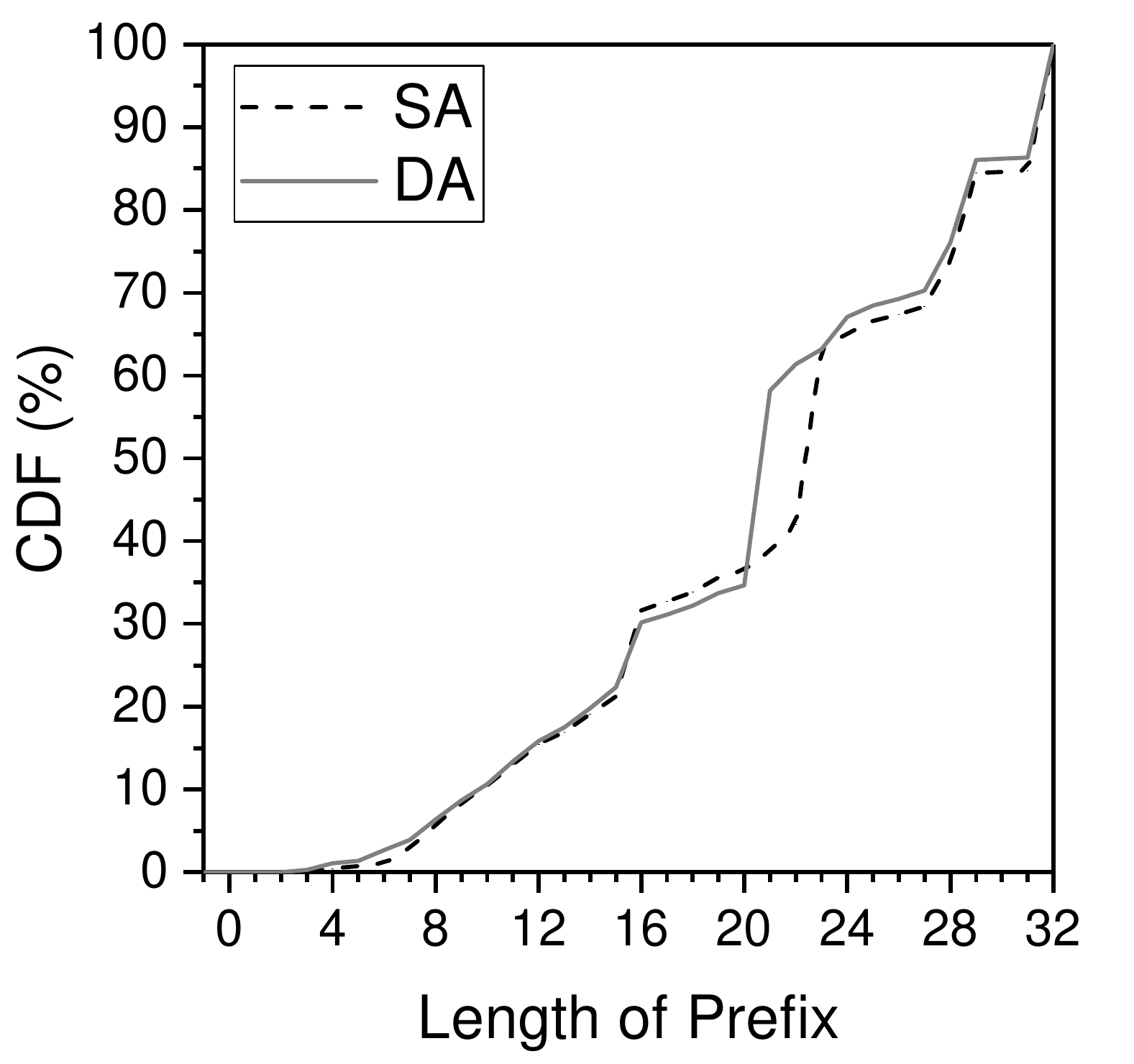}
    \label{fig:cdf:real}
  }
  \caption{Cumulative distribution function of the prefix length of the rules.}
  \label{fig:cdf}
\end{figure}

In our construction policy, we partition the overall length range for prefixes along each dimension separately. Taking SA and DA fields as an example, we partition the range-vector first along the SA dimension, then along the DA dimension. To partition the range-vector along a specific dimension, we first compute the empirical cumulative distribution function (CDF) of the prefix of rules along this dimension (see Figure~\ref{fig:cdf}), and then take the following steps:

\begin{enumerate}
  \item \textbf{Locate partition point:} find the prefix length that most rules are projected to. To this end, we compute the derivative of CDF at each possible prefix length, and then pick out the ones whose derivatives are greater than the average derivative value calculated as the slop of the line connecting the beginning and end points. These values correspond to the prefix lengths that many rules have. In the example of Figure~\ref{fig:cdf:acl}, these values are $12$, $14$, $15$, $16$, $17$, $23$, $24$, $25$, $26$, $30$, $31$, and $32$.

  \item \textbf{Combine length:} Combine adjacent lengths obtained from the first step as a small range, and add the smallest length as a range to incorporate all possible prefix lengths. In this example, we get $[12, 12]$, $[14, 17]$, $[23, 26]$, $[30, 32]$, and we also add the smallest length range $[0, 0]$ as considering the rules whose prefix length is less than $12$.

  \item \textbf{Merge ranges:} Merge two adjacent ranges, if the gap between them does not exceed $D$ while the size of the merged range  is less than $S$ to constrain the probability of rule overlapping. In our example, we set $D=2$ to ensure the two ranges to be close as possible and $S=8$ to ensure the number of overlaps among rules is small. We introduce more criteria for the setup of these parameters in Section~\ref{sect:discu}. With these parameters, we merge $[12, 12]$ and $[14, 17]$ to form a new range $[12, 17]$, and get $[0, 0]$, $[12, 17]$, $[23, 26]$, and $[30, 32]$.

  \item \textbf{Align range:} Align all ranges backwards to ensure the whole range space is included for the rule matching. We get $[0, 11]$, $[12, 22]$, $[23, 29]$, and $[30, 32]$ in our example. These ranges are the partitions along this dimension.
\end{enumerate}
Partition along other dimensions follows the same procedure. Finally, the Cartesian products of every dimension's ranges generate the range-vectors.

\subsection{Stability of Rule Distribution}
\label{sect:part:stab}

In general, for a classifier, its prefix length distribution for each field is stable~\cite{taylor2007classbench}, which barely changes as the rule updates.
We have tracked the changes of the rule distribution in two real ruleset Cloud1 and Cloud2 (Table~\ref{tab:ribs}) during a period of three months that they served for the public network. The result tells us that, although rules have undergone a lot of updates, their prefix length distribution for each field has not changed.

Therefore, for a specific ruleset, we only need to perform the range-vector partition during the initialization period. Of course, if the prefix length distribution of any dimension changes, the partition will be dynamically adjusted. In this case, only a few corresponding hash tables need to be re-hashed, rather than reconstructing the whole classifier. 

%% file: sect-exper.tex

\section{Experimental Results}
\label{sect:exper}

In this section, we perform extensive experiments to compare the performance of RVH with three state-of-the-art packet classification approaches, TSS~\cite{srinivasan1999packet}, PS~\cite{yingchareonthawornchai2016sorted} and TM~\cite{Daly2017TupleMerge}. We evaluate the performance of packet classification, performance of rule updating, and the memory footprint.

\subsection{Platform Implementation}
\label{sect:exper:plat}
Open vSwitch (OVS)~\cite{pfaff2015design} is a software implementation of a distributed virtual multi-layer switch. OVS can support multiple protocols (such as OpenFlow~\cite{mckeown2008openflow}) and standards used in computer networks.
As a widely used software-based switch, OVS is designed with \textit{mega-flow} table and \textit{micro-flow} table to cache rules that were hit recently. This caching mechanism effectively improves the processing speed of packets.
To comprehensively evaluate the performance of different classification schemes, we implemented RVH, TSS, PS and TM to generate the `big' flow table in OVS without changing the caching mechanism. When a packet arrives, it is first matched against rules in the micro-flow table. If it is a miss, the packet is further matched against rules in the mega-flow table, and finally the `big' flow table if it misses again in the match of mega-flow table.

\subsection{Experimental Setup}
\label{sect:exper:setup}
All the experiments were run on a \textit{Sugon} I620-G20 server with an \textit{Intel Xeon} CPU E5-2630 v3 @ 2.40GHz, 32 cores, and 128GB DDR3 memory. Each core is integrated with a 64KB L1 data cache and a 256KB L2 cache. A 20MB L3 cache is shared among all cores. \textit{Ubuntu} 16.04.1 with \textit{Linux kernel} 4.10.0 is installed as the operating system.

\begin{table}[!t]
  \centering
  \caption{Rulesets for Experiments}
  \label{tab:ribs}
  \begin{tabular}{ccc}
    \toprule
    Rulesets & \# of Rules & \# of Range-vectors\\
    \midrule
    ACL1   & 95399  & 16 \\
    ACL2   & 93912  & 16 \\
    FW1    & 215210 & 8  \\
    FW2    & 209185 & 8  \\
    IPC1   & 29078  & 4  \\
    IPC2   & 31976  & 4  \\
    Cloud1 & 1427   & 24 \\
    Cloud2 & 16603  & 24 \\
    \bottomrule
  \end{tabular}
\end{table}
To evaluate the performance of packet classification algorithms, we used 8 different rulesets, which fall into four types (ACL, FW, IPC and Cloud) as shown in Table~\ref{tab:ribs}. The first six rulesets are generated by ClassBench~\cite{taylor2007classbench}.
ClassBench generates packet traces and rules following the distribution of real rulesets. The last two rulesets were collected from two operating OpenStack cloud nodes of a major ISP\footnote{Due to confidential agreement, we are not allowed to reveal the name of the company.}.

In the experiments, we take SA and DA as the fields to match, because other fields are very sparse. We have shown the distribution of rules in four types of rulesets in Figure~\ref{fig:dist}. The distribution is not uniform, but clustered in some ranges, especially for IPC and FW rulesets.

\subsection{Updating}
\label{sect:exper:update}

We first examine the performance for updating rulesets, because supporting fast rule update is mandatory in SDN and cloud networks. For instance, OpenFlow rules are often dynamically changed by controllers in SDN networks. For this reason, TSS, which was proposed about 20 years ago for fast update, is used by OVS.

\begin{figure}[!t]
  \centering
  \subfigure[with different rulesets]{
    \includegraphics[width=0.7\linewidth]{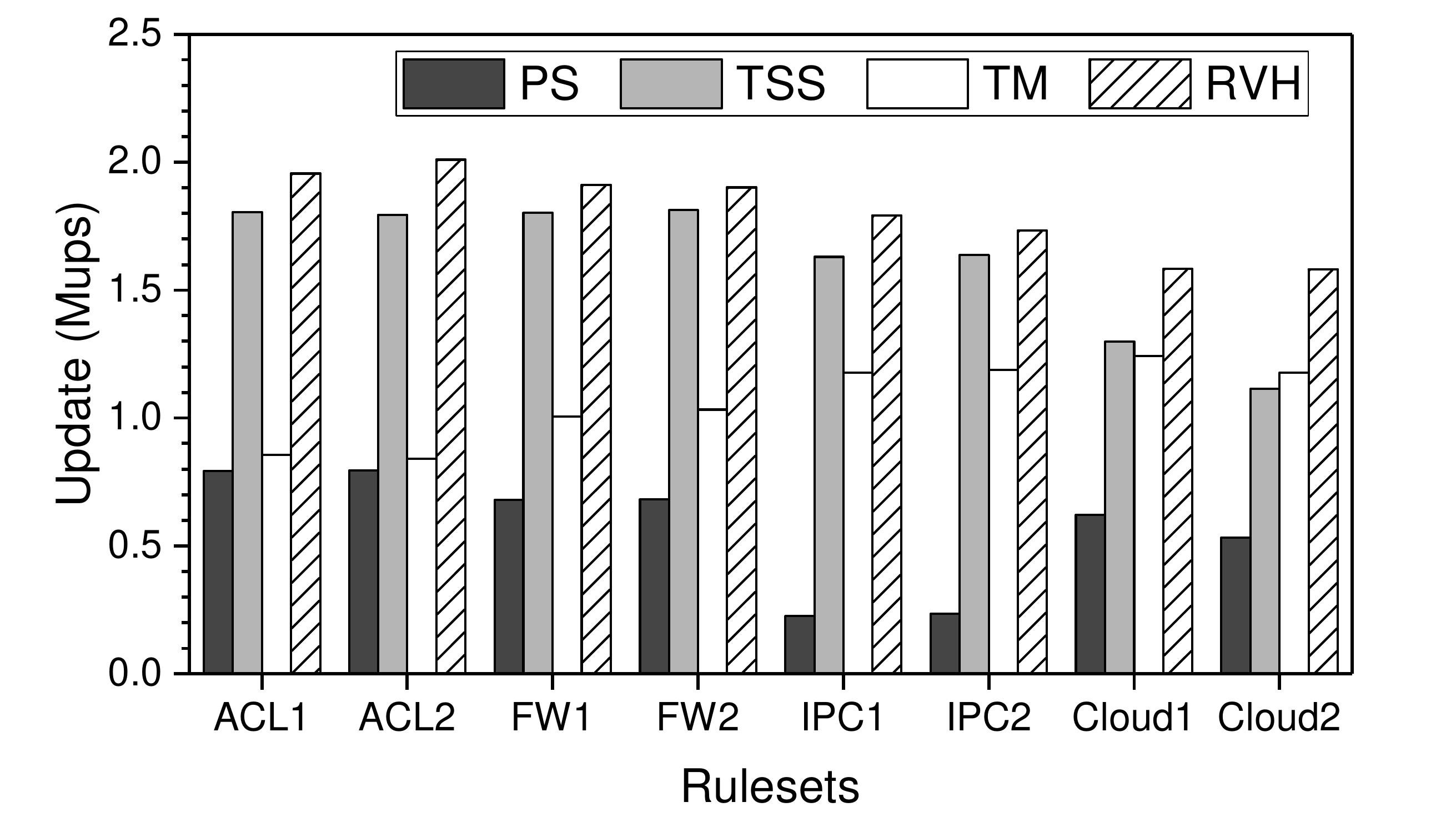}
    \label{fig:update:data}
  }
  \subfigure[with different size of classifiers]{
    \includegraphics[width=0.7\linewidth]{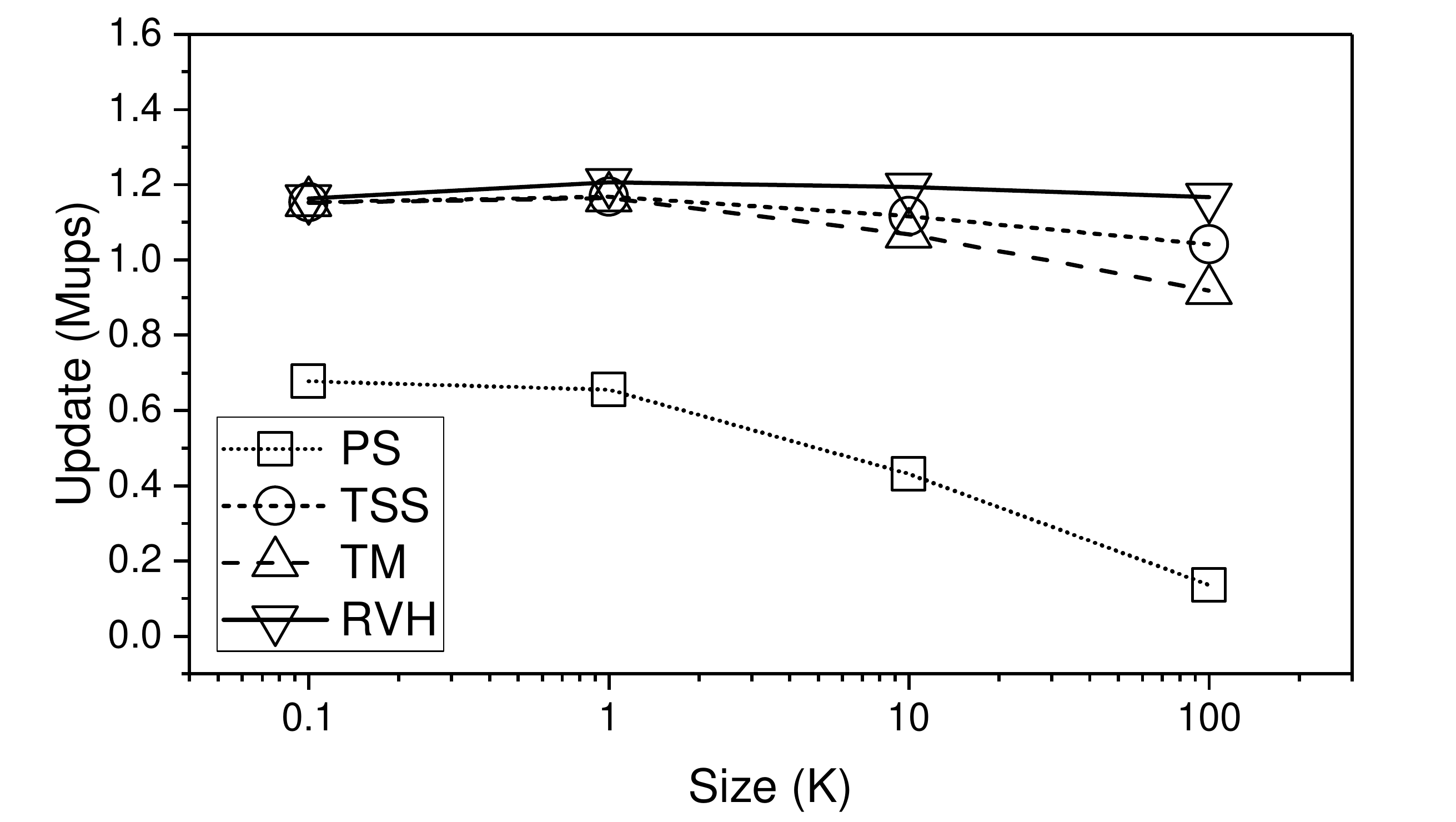}
    \label{fig:update:size}
  }
  \caption{Comparisons of updating performance: (a) varying rulesets; (b) varying the size of classifiers.}
  \label{fig:update}
\end{figure}
We consider both insertion and deletion operations. Each ruleset was first divided into five subsets evenly, where four were used as the original ruleset to initialize the classifier, and the remaining one was inserted into the classifier thereafter.
We use \textit{million updates per second} (Mups) to measure the updating speed. As shown in Figure~\ref{fig:update:data}, RVH outperforms other three approaches greatly. The updating speed of RVH is $3.9\times$ that of PS on average, $1.7\times$ that of TM and $1.1\times$ that of TSS on average. While the time complexity of TSS is the same as that of RVH (see Table~\ref{tab:complex}), RVH achieves a slightly higher performance. This is because RVH reduces the number of hash tables, which causes the cache hit rate to be higher than that of TSS. The low updating performance of PS is due to its use of a tree structure, which degrades the performance. Besides, PS's updating performance is not stable across rulesets, because the decision tree is dependent on the properties of rulesets. TM fails to achieve a high updating performance as RVH or TSS because of the ruleset overlapping and classifier reconstruction.

We then evaluate the impact of classifier size (i.e., the number of rules in the classifier) on the updating performance in Figure~\ref{fig:update:size}. In this set of experiments, we used the cloud ruleset and formed it into four parts with various sizes: 0.1K, 1K, 10K and 100K. As expected, the updating speed decreases with the increase of classifier size. Nevertheless, the speed of RVH decreases much slower than the other approaches.

\subsection{Packet Classification without Updating}
\label{sect:exper:classwo}

\begin{figure}[!t]
  \centering
  \subfigure[with different rulesets]{
    \includegraphics[width=0.7\linewidth]{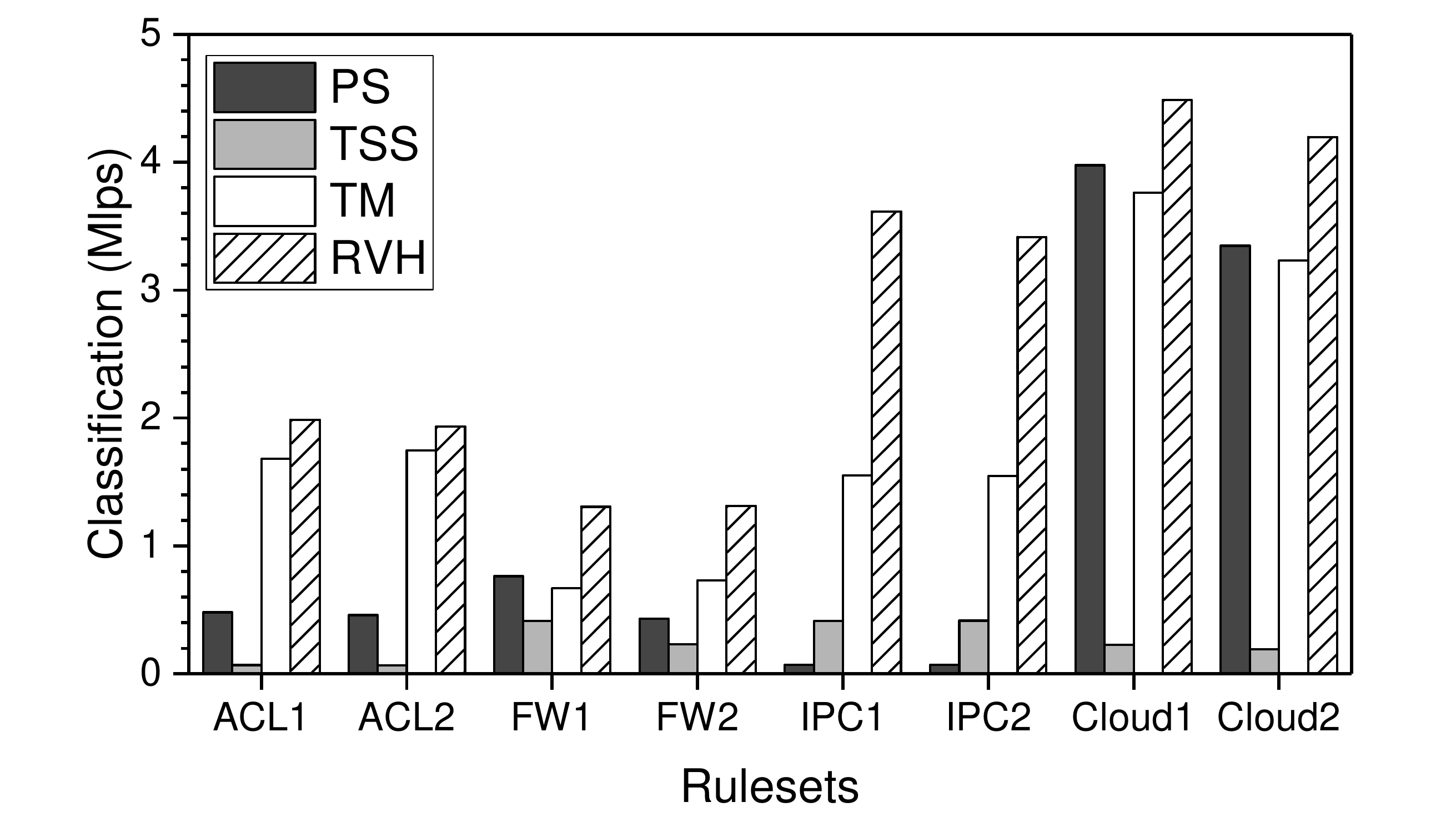}
    \label{fig:lookup:data}
  }
  \subfigure[with different size of classifiers]{
    \includegraphics[width=0.7\linewidth]{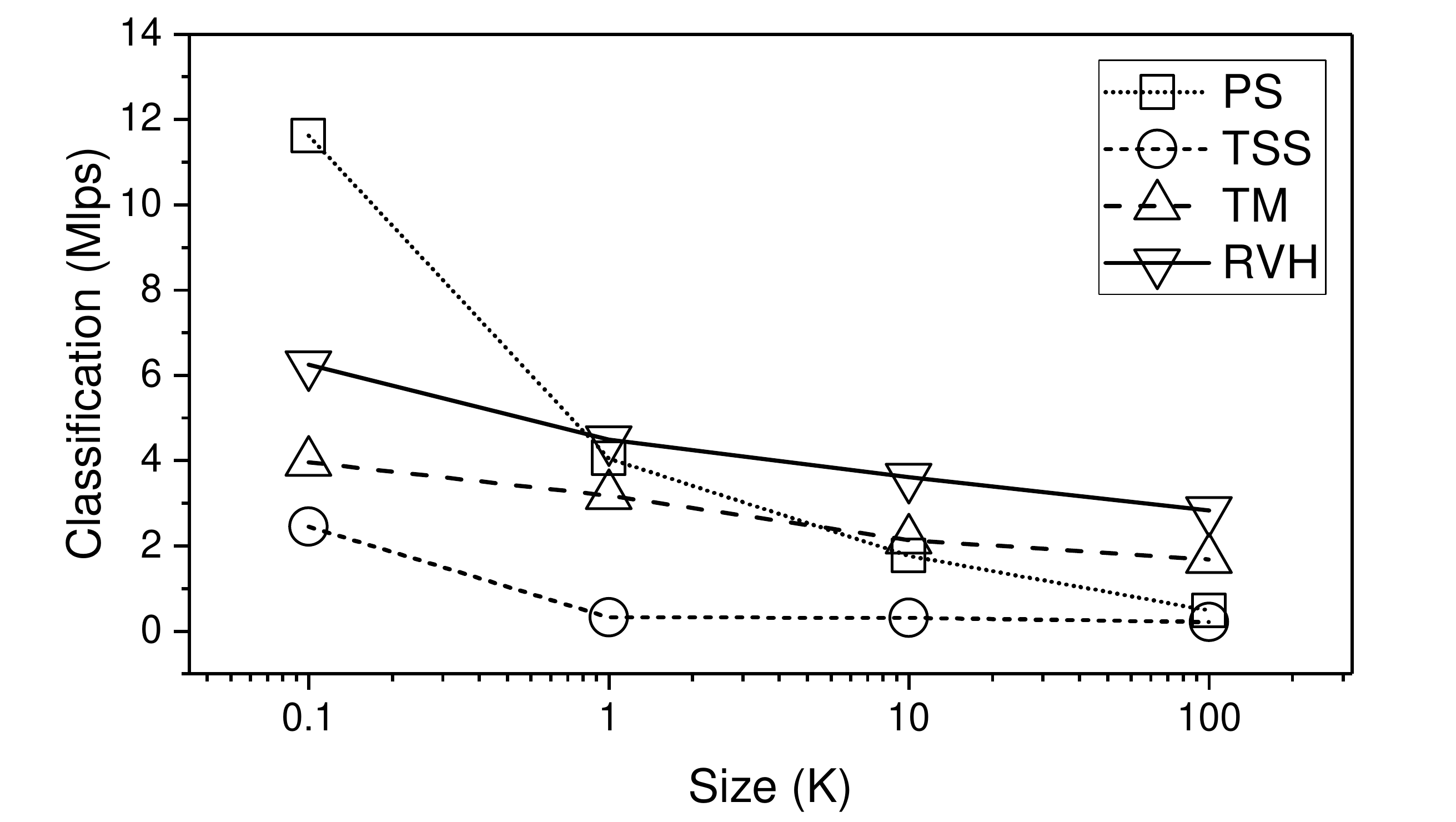}
    \label{fig:lookup:size}
  }
  \caption{Packet classification performance without updating: (a) varying rulesets; (b) varying the size of classifiers.}
  \label{fig:lookup}
\end{figure}

Next, we examine the packet classification performance when there are no updating operations in Figure~\ref{fig:lookup}, where we use \textit{million lookups per second} (Mlps) as the unit.

We can observe that the packet classification performance of RVH is $15.7\times$ that of TSS, $14.1\times$ that of PS and $1.6\times$ that of TM on average. By using range-vectors instead of tuples in TSS, RVH significantly reduces the number of hash tables, leading to superior performance. PS was proposed to improve the packet classification performance of TSS, but as we can see, for some rulesets (e.g., IPC), its performance is even lower than TSS. This again shows that the performance of decision tree (used by PS) depends on the properties of rulesets. TM achieves a lower performance than RVH, because it maintains a slightly larger number of hash tables and also suffers from the issue of rule overlapping.

While varying the size of classifier (see Figure~\ref{fig:lookup:size}), we observe that RVH outperforms others when the sizes of rulesets exceed 1K rules. Below this value, PS has a better performance. Nevertheless, the performance of PS degrades sharply when increasing the size.

\subsection{Packet Classification with Updating}
\label{sect:exper:classwi}
\begin{figure*}[!t]
  \centering
  \subfigure[ACL1]{
    \includegraphics[width=0.22\linewidth]{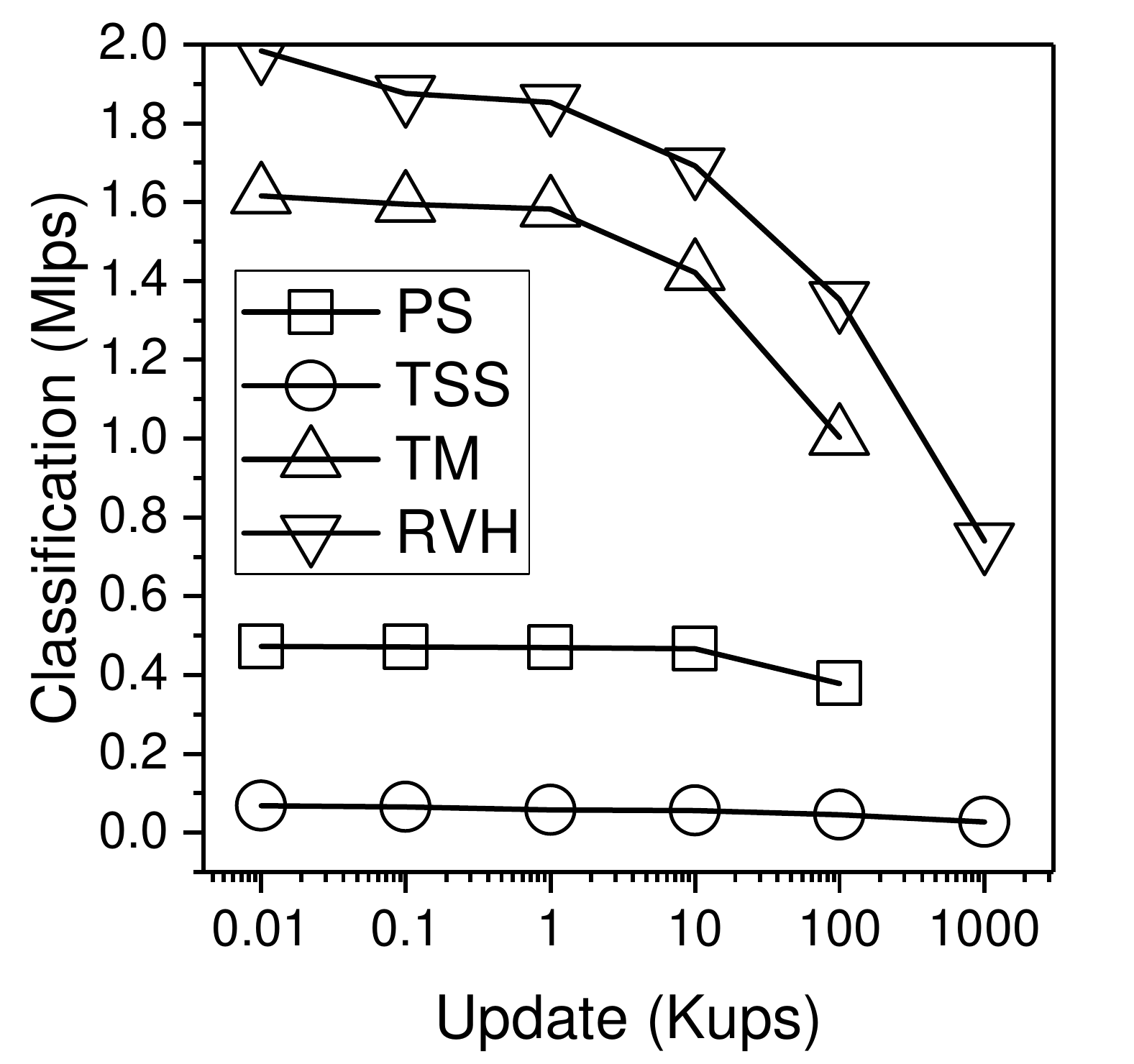}
    \label{fig:sim:acl1}
  }
  \subfigure[ACL2]{
    \includegraphics[width=0.22\linewidth]{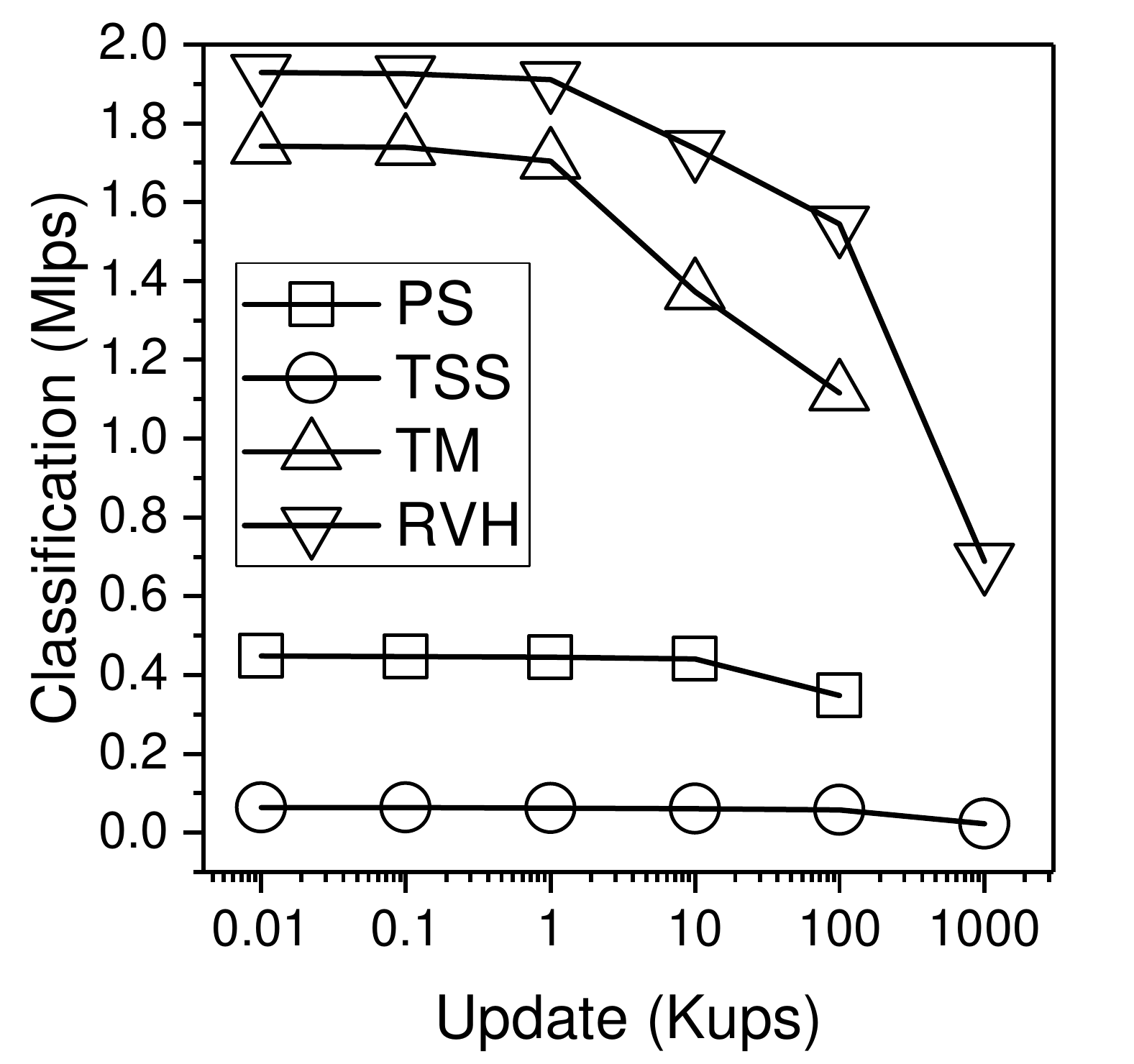}
    \label{fig:sim:acl2}
  }
  \subfigure[FW1]{
    \includegraphics[width=0.22\linewidth]{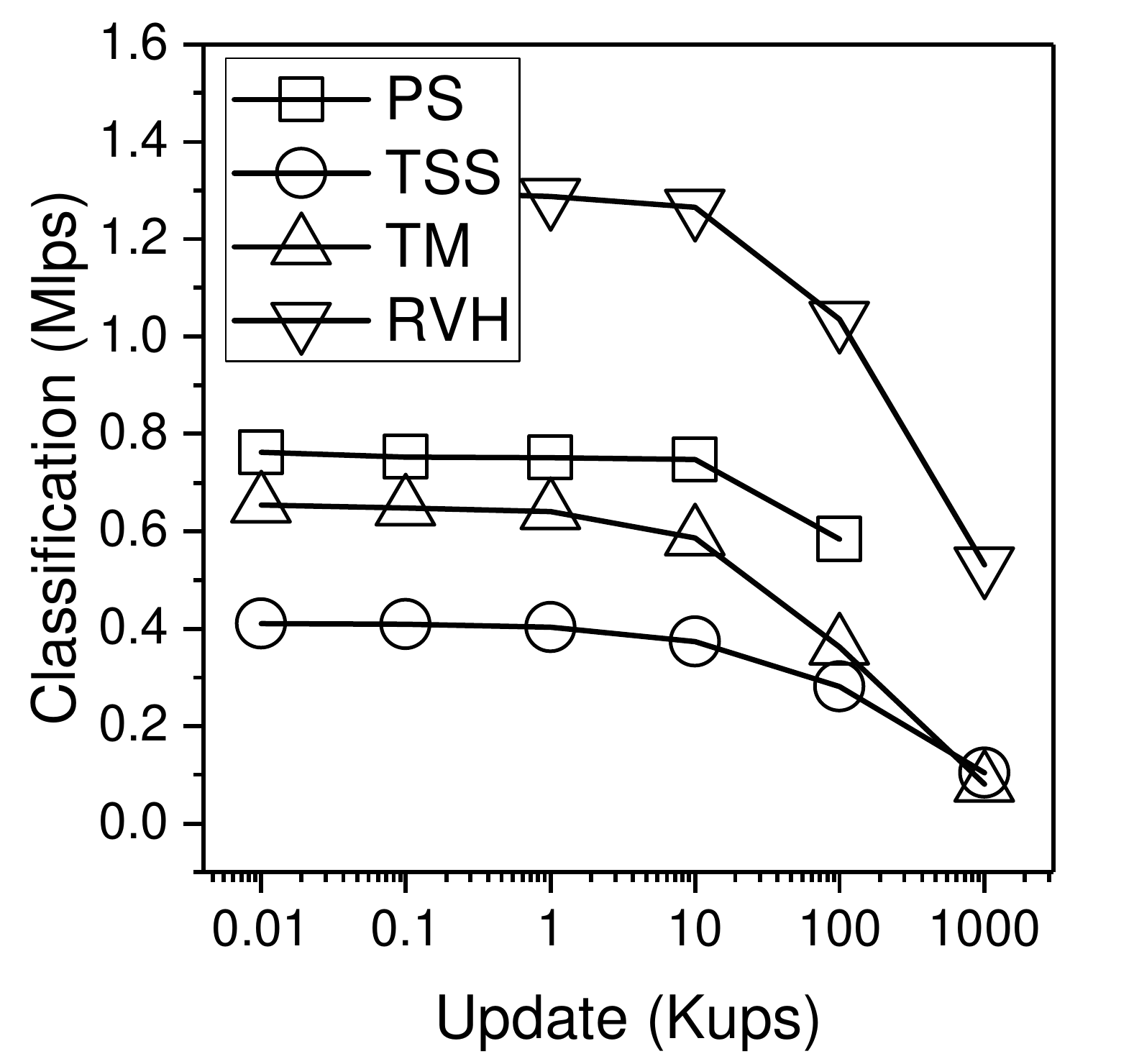}
    \label{fig:sim:fw1}
  }
  \subfigure[FW2]{
    \includegraphics[width=0.22\linewidth]{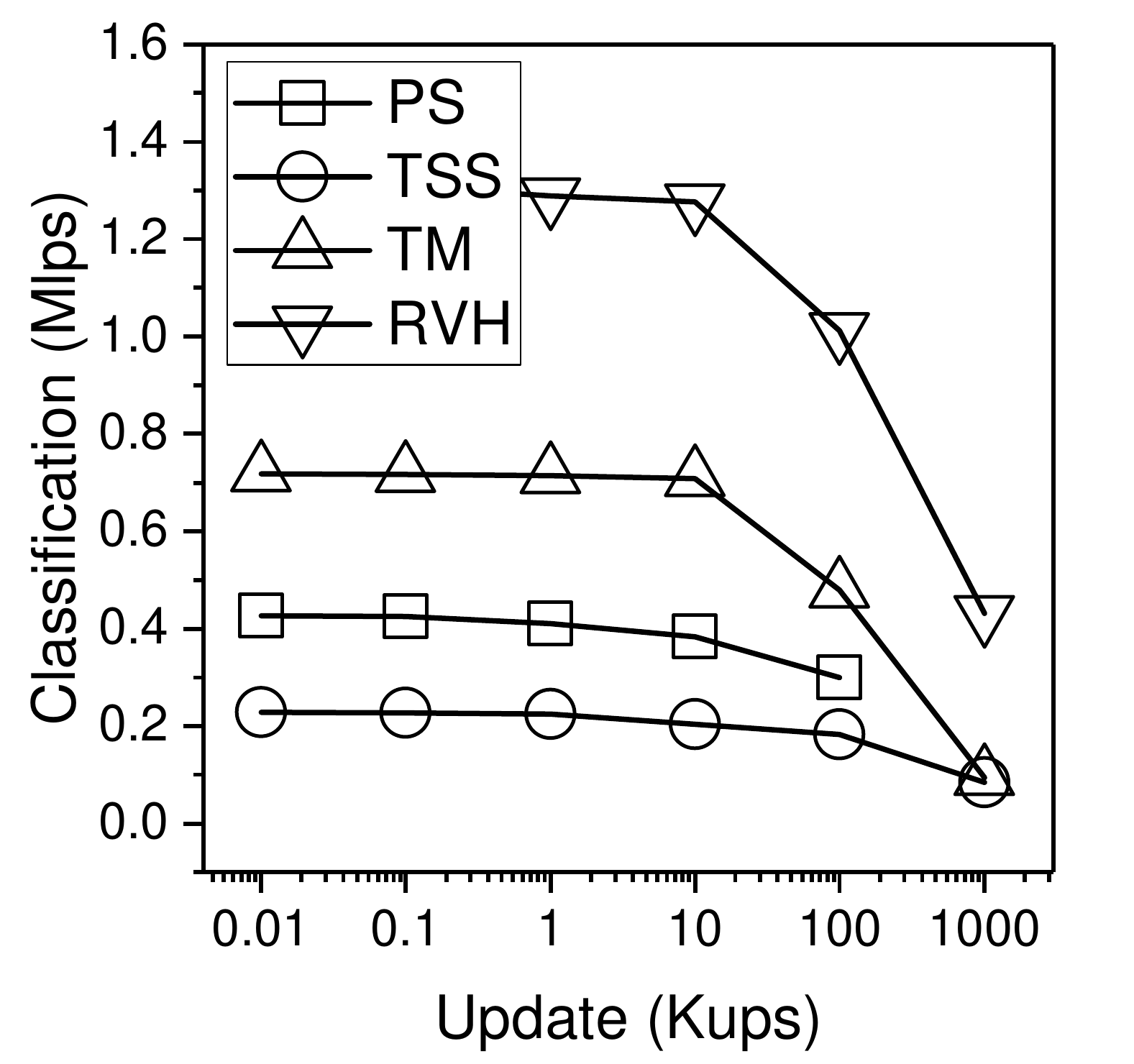}
    \label{fig:sim:fw2}
  }
  \subfigure[IPC1]{
    \includegraphics[width=0.22\linewidth]{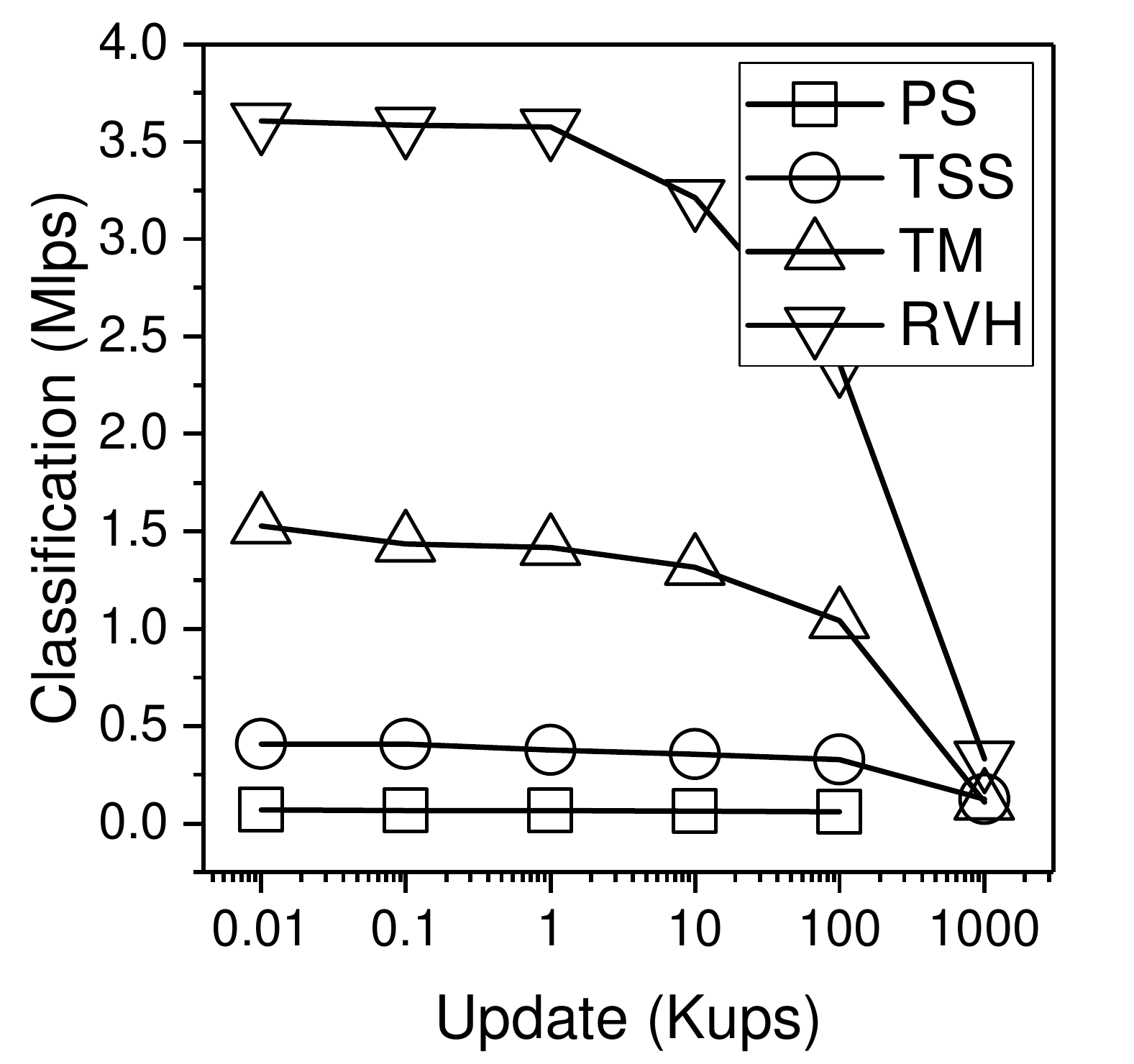}
    \label{fig:sim:ipc1}
  }
  \subfigure[IPC2]{
    \includegraphics[width=0.22\linewidth]{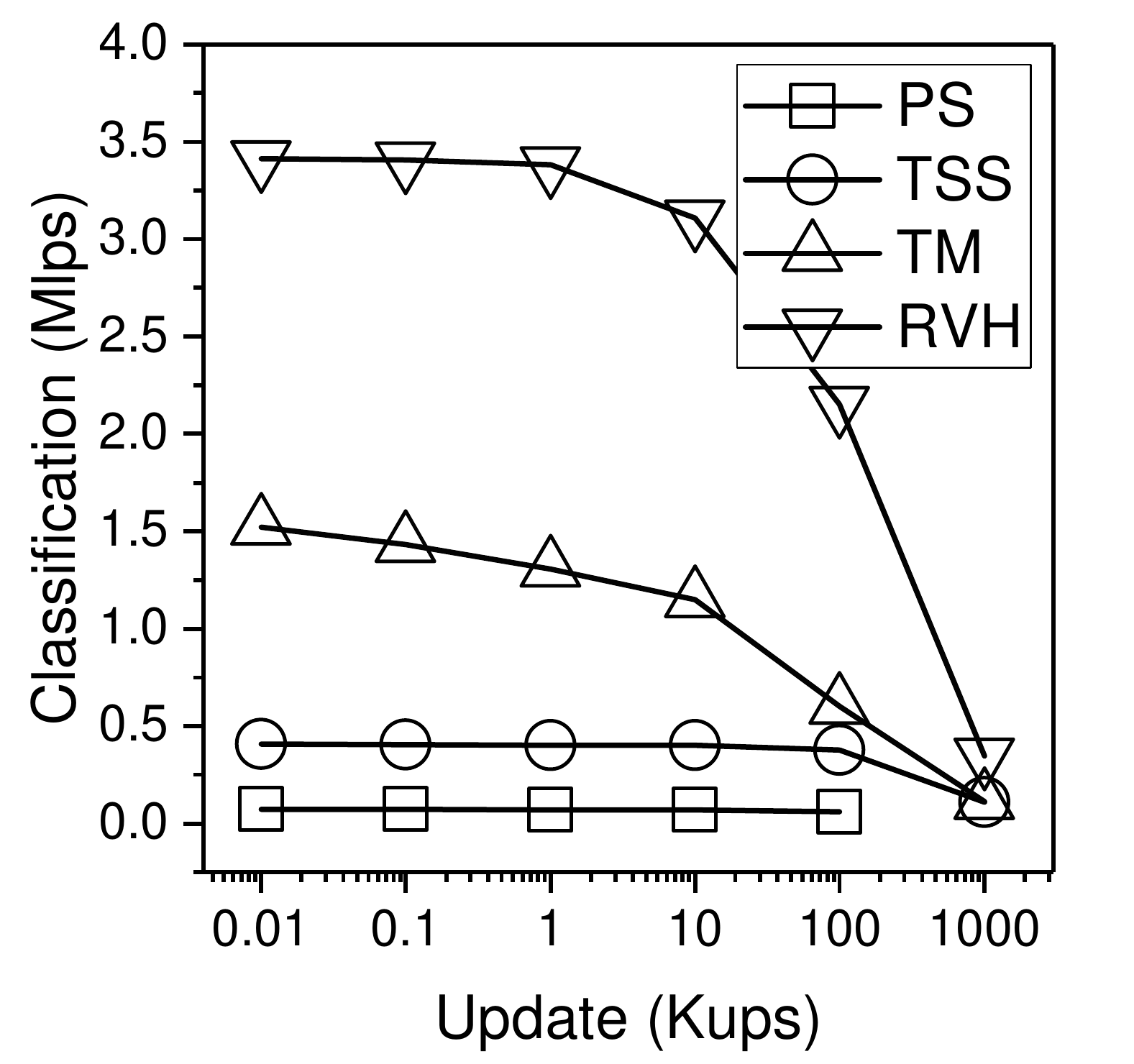}
    \label{fig:sim:ipc2}
  }
  \subfigure[Cloud1]{
    \includegraphics[width=0.22\linewidth]{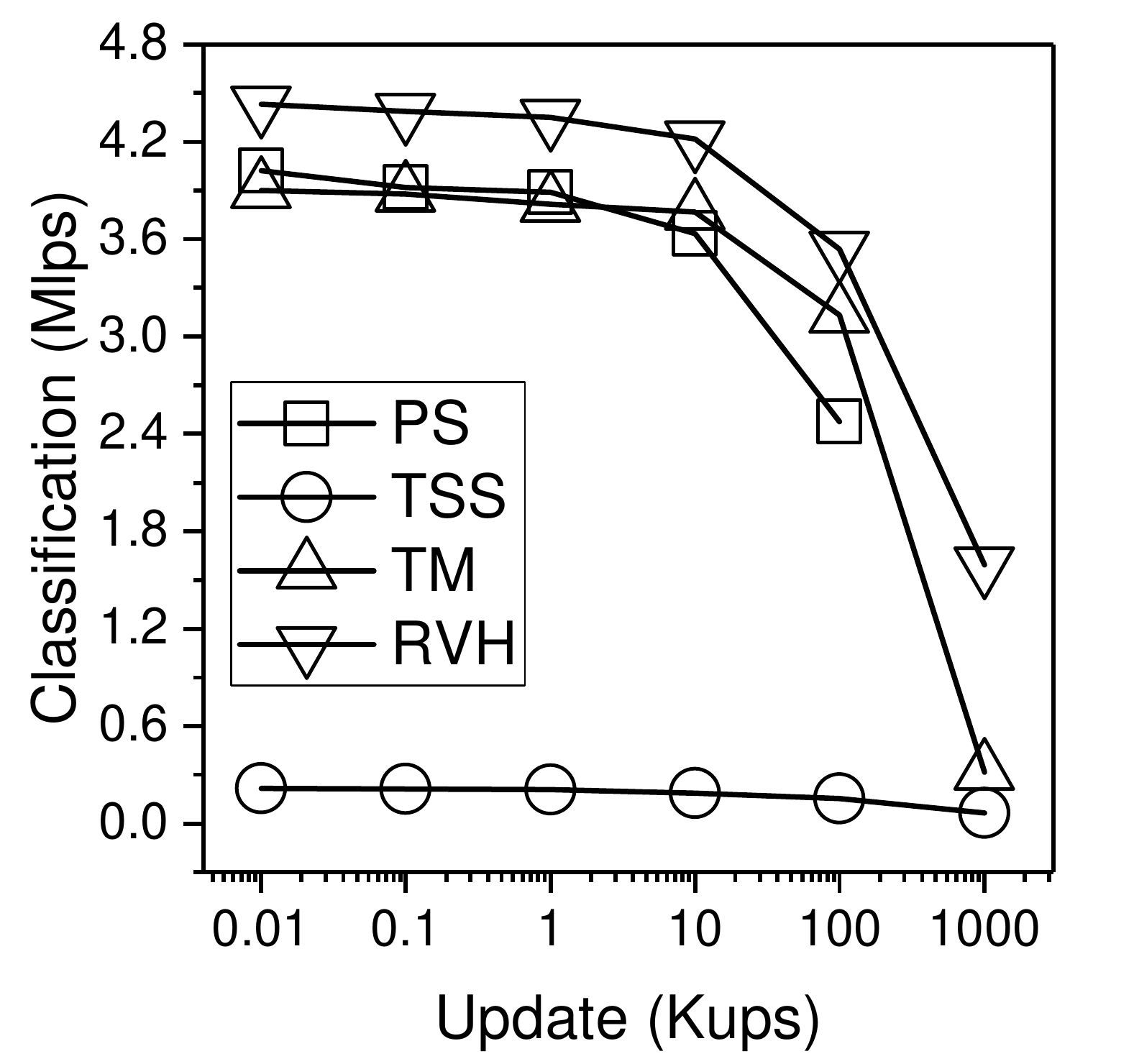}
    \label{fig:sim:real1}
  }
  \subfigure[Cloud2]{
    \includegraphics[width=0.22\linewidth]{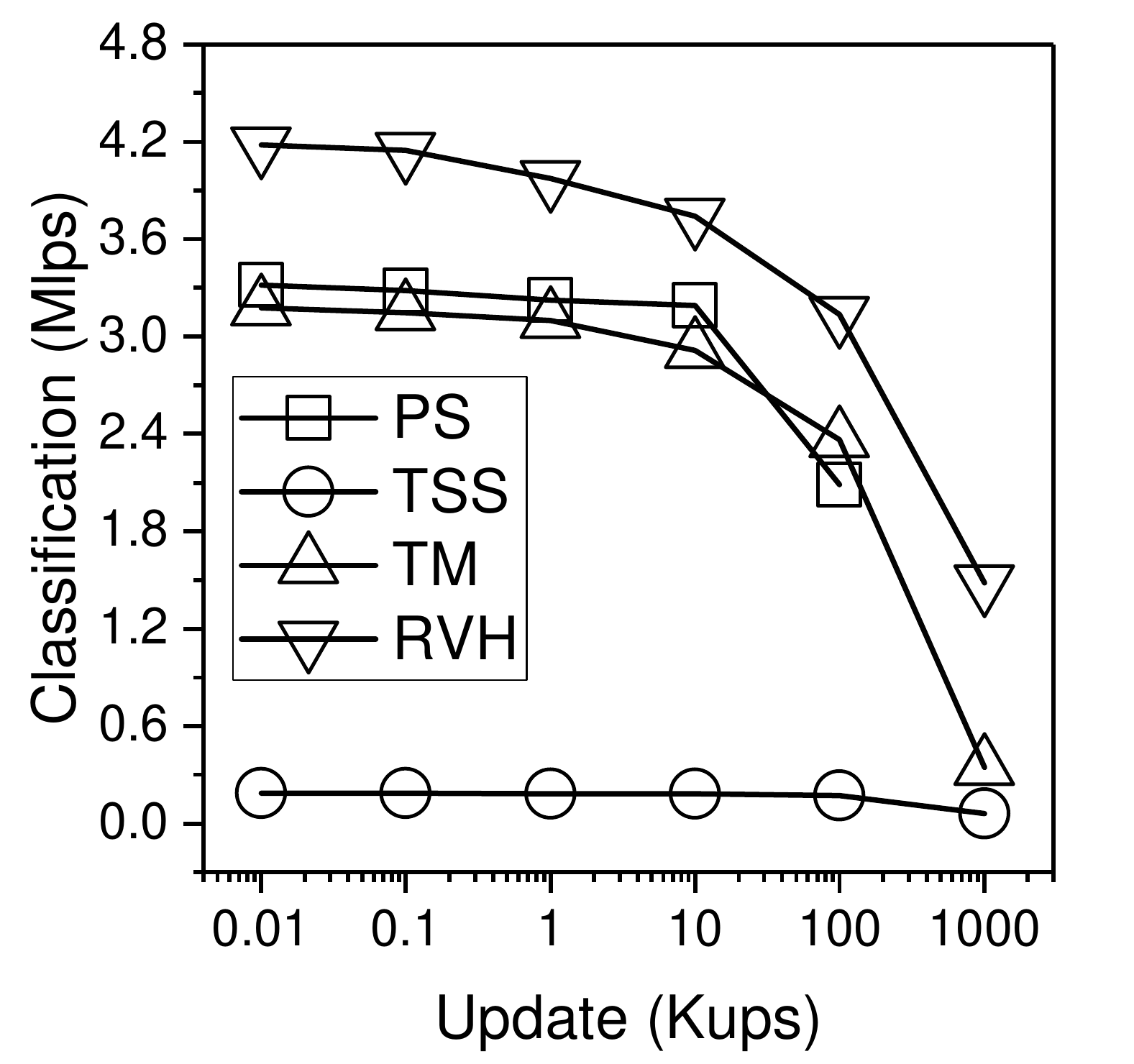}
    \label{fig:sim:real2}
  }
  \caption{Classification  with different update speeds. }
  \label{fig:sim}
\end{figure*}

We then evaluate the impact of classifier updating on packet classification performance using the 8 rulesets in Figure~\ref{fig:sim}, where we vary the update speed. We make the following interesting observations. First, RVH achieves the highest performance regardless of rulesets and the updating speed. Second, very frequent updates will degrade the packet classification performance, especially when the update speed exceeds 10Kups. Third, when the update speed reaches 1000kups, except for RVH, the other three approaches (particularly PS) almost cannot classify packets. Fourth, PS's performance is heavily dependent on ruleset properties. For instance, PS achieves the performance comparable to TM for cloud rulesets, while it has the lowest performance for IPC rulesets.

\subsection{Memory Footprint}
\label{sect:exper:mem}

\begin{figure}[!t]
  \centering
  \subfigure[with different rulesets]{
    \includegraphics[width=0.7\linewidth]{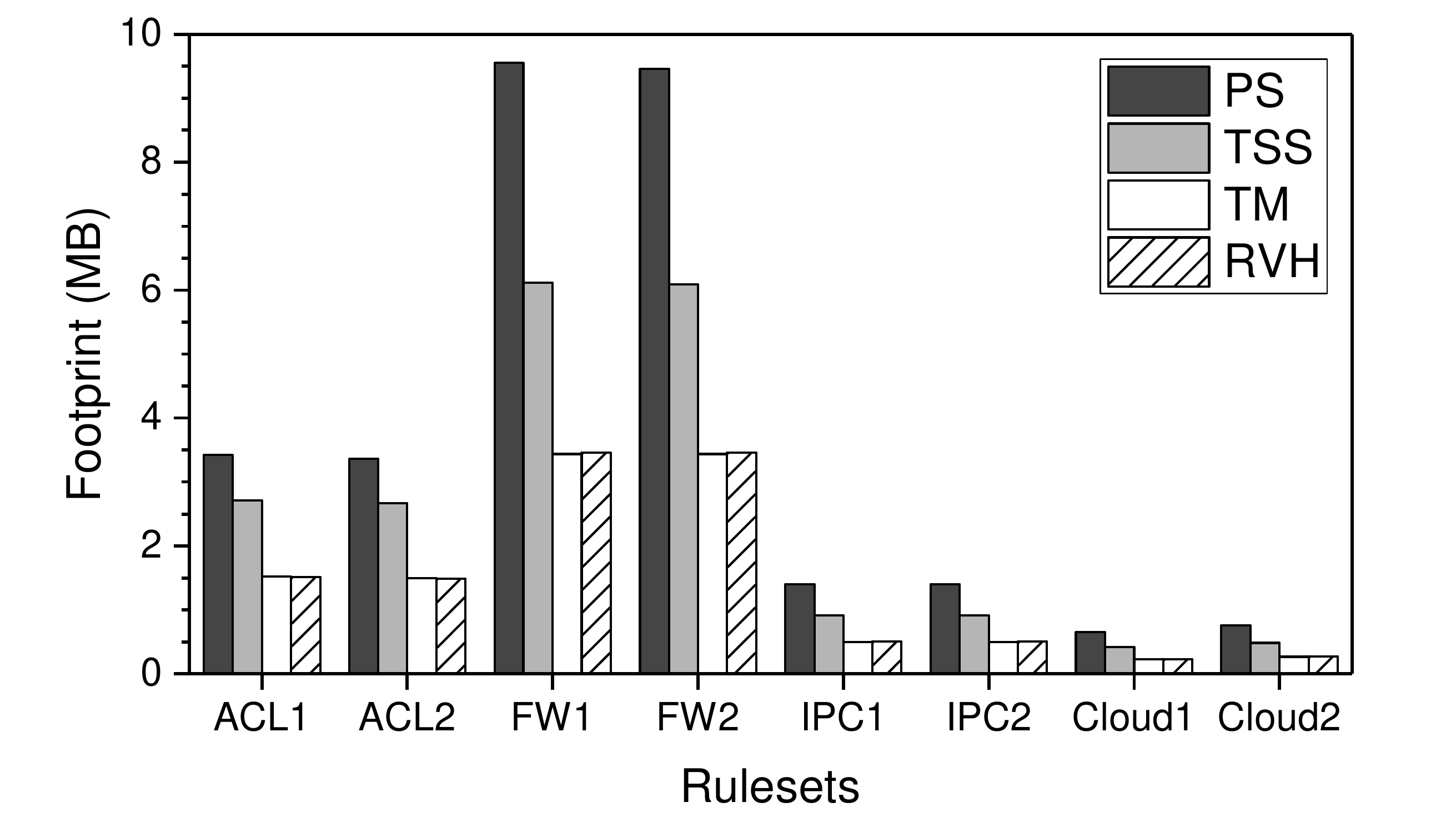}
    \label{fig:memory:data}
  }
  \subfigure[with different size of classifiers]{
    \includegraphics[width=0.7\linewidth]{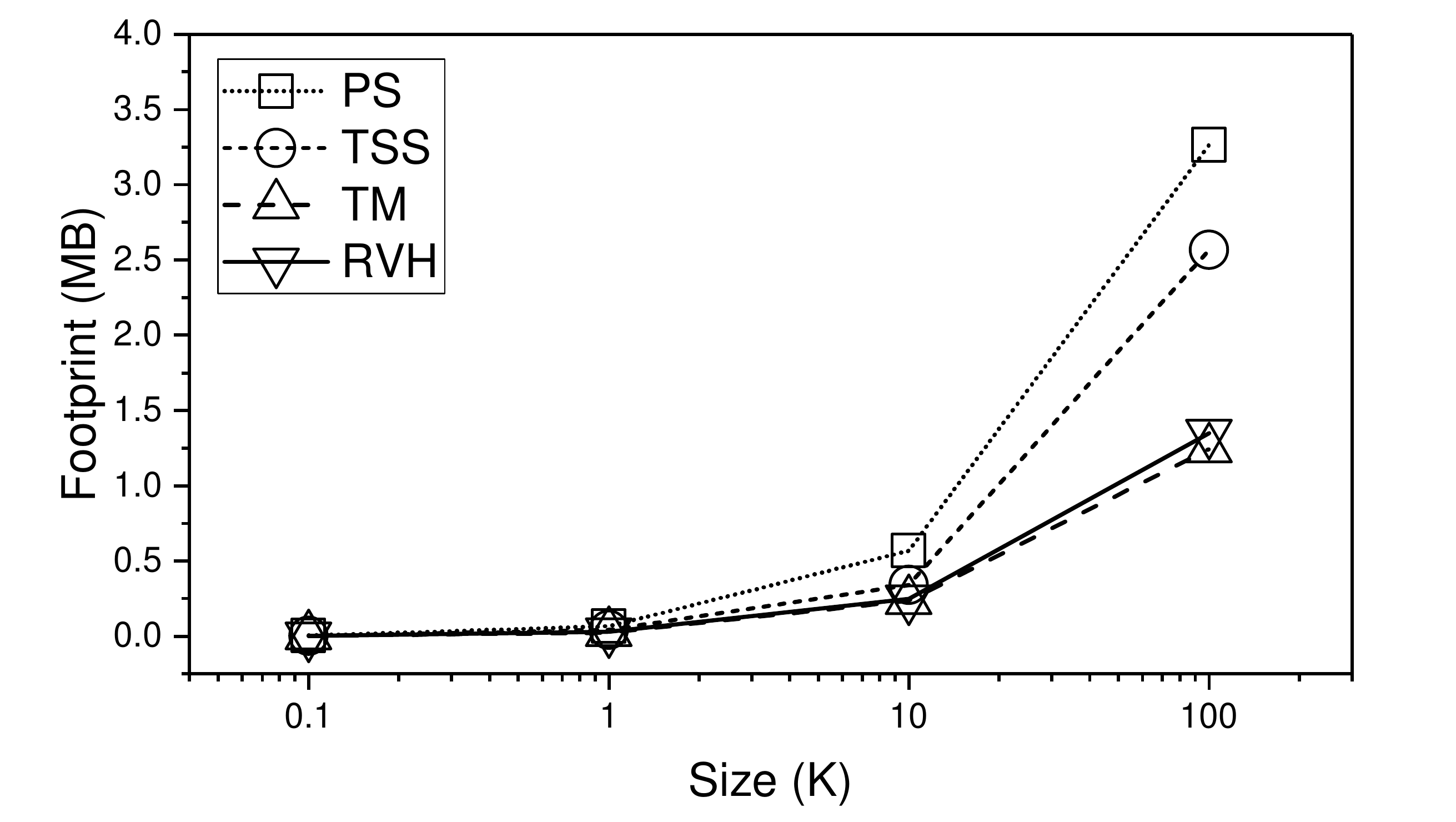}
    \label{fig:memory:size}
  }
  \caption{Memory footprint: (a) varying rulesets; (b) varying the sizes of classifiers.}
  \label{fig:memory}
\end{figure}
Finally, we examine the memory footprint of different approaches in Figure~\ref{fig:memory:data}. We again see that RVH outperforms others in terms of memory footprint as well. Indeed, RVH and TM are very similar in terms of memory footprint. The memory footprint of RVH is only $38\%$ and $56\%$ of that used by PS and TSS respectively. Recall that PS uses decision tree structures to filter tuples. As a decision tree takes a lot of memory, the footprint of PS is significantly higher than those of other two approaches. Cuckoo hash~\cite{Pagh2004Cuckoo} used in RVH and TSS can greatly improve the utilization of hash tables. 

From Figure~\ref{fig:memory:size}, we can see that as the classifier becomes larger, the memory footprint of each algorithm also increases. Nevertheless, the memory footprint of RVH and TM are always less than those of TSS and PS.

%% file: sect-discu.tex
\section{Discussion}
\label{sect:discu}

In this section, we discuss some potential steps to take to further improve the performance of RVH in our future work.

Our algorithm partitions the overall prefix length into a few fine-grained range-vectors based on the distribution of rules. As a result, classifying a packet only requires a few hashes, and the performance is significantly improved. This sounds like an ad hoc, but in fact RVH can be applied to any classifier and ruleset.

To further improve the classification speed, we also introduce two procedures, range-vector priority sorting and overlapped rule priority sorting.
However, it takes extra time to select the rule for match when multiple rules are found to be related to a hash key.
\begin{figure}[!t]
  \centering
  \includegraphics[width=0.7\linewidth]{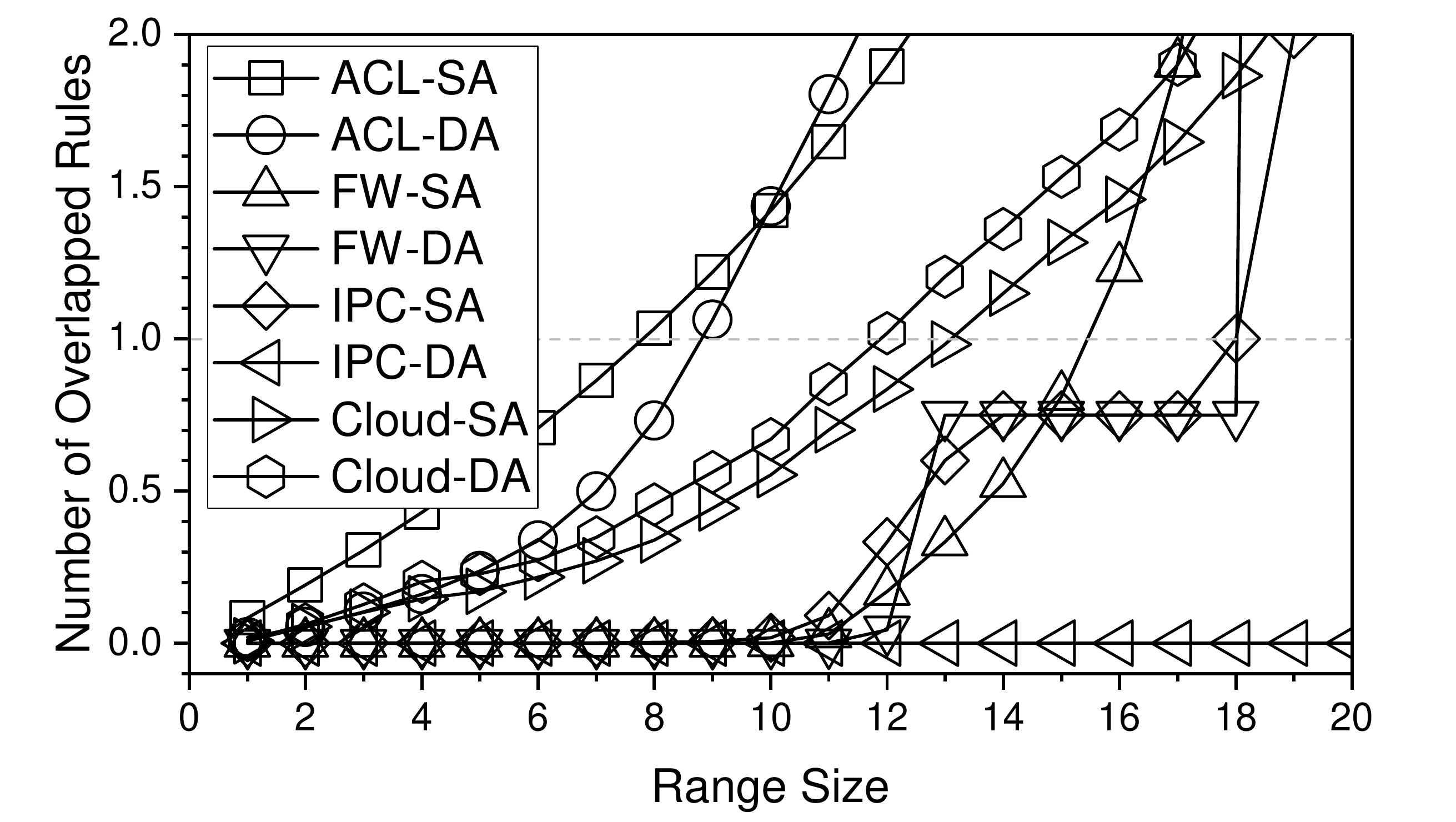}
  \caption{Average number of overlapped rules for each entry, varying by the size of prefix length ranges.}
  \label{fig:overlap}
\end{figure}
To reduce the time for rule verification, in Figure~\ref{fig:overlap}, we show statistics on the average number of overlapped rules for each entry based on the sizes of prefix-length ranges. We take SA and DA as example fields to match in this study.
The number of overlapped rules for each entry grows rapidly as the size of range increases. We found that as long as the range size is less than $8$, the average number of overlapped rules for each entry is less than $1.0$. Thus in our partition policy (Section~\ref{sect:part:policy}), we would ensure that the size of each range is less than $8$.

In the future, we will extend RVH to run in multi-core systems.
There are two modes in parallel, \textit{Run to Complete} (RTC) and \textit{Software Pipline} (SPL). We intend to combine these two modes.
For packet classification, we intend to run an independent thread for each core while sharing the ruleset for each process.
For rule update, we will separate the two operations, locating and updating, and put them into a pipeline.
The multi-thread RVH just needs to lock the corresponding rules not the whole classifier, when updating rules. We expect these procedures can significantly increase the speeds for both packet classification and rule update. 

%% file: sect-conclu.tex
\section{Conclusion}
\label{sect:conclu}

In this paper, we propose RVH, a packet classification approach that supports both fast rule update and fast packet classification. RVH is built on the key observation of biased distribution of prefix lengths over the multi-dimensional space. Specially, RVH divides the whole range-length of prefixes into a few range-vectors to group rules into a small number of hash tables, which significantly increases the packet classification speed while ensuring fast rule update. We provide details on the design of RVH and introduce additional procedures to take for further speeding up the classification. We have evaluated the performance of RVH using both the rulesets generated by ClassBench and the rulesets in production cloud networks. The results consistently show the superior performance of RVH in both rule update and packet classification, compared with the state-of-the-art approaches.

%% file: sect-acks.tex

\section*{Acknowledgment}
This work is supported by the Major State Basic Research Development Program of China~(973 Program) under Grant No.~2012CB315805 and the National Natural Science Foundation of China under Grant Nos.~61472130 and~61473123.